\documentclass{sig-alternate}
\usepackage{amsmath,amsfonts,amssymb}
\usepackage{times}
\usepackage[usenames,dvipsnames]{xcolor}
\usepackage{pgfplots}
\usepackage{url}
\usepackage{caption}
\usepackage{verbatim}
\usepackage{hyperref}
\usepackage{epsfig}
\usepackage{latexsym}
\usepackage{MnSymbol} 
\usepackage{multirow}
\usepackage{subfigure}
\usepackage{graphicx}
\usepackage{verbatim}
\usepackage{framed}
\usepackage{enumitem}
\usepackage[nocompress]{cite}
\usepackage{relsize}
\usepackage{mathtools}
\usepackage{balance}
\usepackage[linesnumbered,ruled,vlined,boxed]{algorithm2e}
\usepackage{grffile}
\usepackage{paralist}
\usepackage{xspace}

\usepackage[hang,flushmargin]{footmisc}

\pgfplotsset{
  width=6cm,
  height=4cm, 
  compat=newest,
  xlabel near ticks,
  ylabel near ticks}

\makeatletter
\def\blfootnote{\xdef\@thefnmark{}\@footnotetext}
\makeatother

\newcommand{\agpneutral}[1]{#1}

\newcommand{\agpcomment}[1]{}

\newcommand{\alkim}[1]{\textcolor{ForestGreen}{Albert: #1}}

\newcommand{\ntail}{{\sc NeedleTail}\xspace}

\newtheorem{definition}{Definition}[section]
\newtheorem{example}[definition]{Example}

\newtheorem{problem}{Problem}
\newtheorem{theorem}[definition]{Theorem}
\newtheorem{corollary}[definition]{Corollary}

\newtheorem{remark}{Remark}
\newtheorem{lemma}{Lemma}

\newcommand{\algorithmfootnote}[2][\footnotesize]{%
  \let\old@algocf@finish\@algocf@finish% Store algorithm finish macro
  \def\@algocf@finish{\old@algocf@finish% Update finish macro to insert "footnote"
    \leavevmode\rlap{\begin{minipage}{\linewidth}
    #1#2
    \end{minipage}}%
  }%
}

\DeclareMathOperator*{\E}{\mathrm{E}\xspace}

\newcommand{\calY}{\mathcal{Y}\xspace}

\newcommand{\squishlist}{
   \begin{list}{$\bullet$}
    { \setlength{\itemsep}{0pt}
      \setlength{\parsep}{2pt}
      \setlength{\topsep}{0pt}
      \setlength{\partopsep}{0pt}
      \leftmargin=25pt
\rightmargin=0pt
\labelsep=5pt
\labelwidth=10pt
\itemindent=0pt
\listparindent=0pt
\itemsep=\parsep
    }
}
\newcommand{\squishend}{\end{list}}

\newenvironment{denselist}{
    \begin{list}{\tiny{$\bullet$}}%
    {\setlength{\itemsep}{0ex} \setlength{\topsep}{0ex}
    \setlength{\parsep}{0pt} \setlength{\itemindent}{0pt}
    \setlength{\leftmargin}{1.5em}
    \setlength{\partopsep}{0pt}}}%
    {\end{list}}

\newcommand{\eat}[1]{}
\newcommand{\papertext}[1]{}
\newcommand{\techreport}[1]{#1}
\newcommand{\ifocus}[0]{{\sc IFocus}\xspace}
\newcommand{\iold}[0]{{\sc IRefine}\xspace}
\newcommand{\ifocusr}[0]{{\sc IFocusR}\xspace}
\newcommand{\ioldr}[0]{{\sc IRefineR}\xspace}

\newcommand{\ir}[0]{{\sc RoundRobin}\xspace}
\newcommand{\irr}[0]{{\sc RoundRobinR}\xspace}
\newcommand{\scan}[0]{{\sc Scan}\xspace}

\newcommand{\techreporttext}[1]{}
\newcommand{\stitle}[1]{\vspace{0.25em}\noindent\textbf{#1}}

\makeatletter
\def\@copyrightspace{\relax}
\makeatother
\begin{document}

\title{Rapid Sampling for Visualizations \\ with Ordering Guarantees}

\numberofauthors{6}
\author{
\alignauthor Albert Kim \\ 
\affaddr{MIT} \\ 
\affaddr{\small ${\tt alkim@csail.mit.edu}$ }
\alignauthor Eric Blais \\ 
\affaddr{MIT and University of Waterloo} \\
\affaddr{\small ${\tt eblais@uwaterloo.ca}$}
\alignauthor Aditya Parameswaran \\ 
\affaddr{MIT and Illinois (UIUC)} \\ 
\affaddr{\small ${\tt adityagp@illinois.edu}$ }
\and
\alignauthor Piotr Indyk \\ 
\affaddr{MIT} \\
\affaddr{\small ${\tt indyk@mit.edu}$}
\alignauthor Sam Madden \\ 
\affaddr{MIT} \\ 
\affaddr{\small ${\tt madden@csail.mit.edu}$ }
\alignauthor Ronitt Rubinfeld \\ 
\affaddr{MIT and Tel Aviv University}\\
\affaddr{\small ${\tt ronitt@csail.mit.edu}$}
}
\maketitle                                                                                       

\begin{abstract}
%!TEX root=main.tex

\agpneutral{Visualizations are frequently used as a means to understand trends and gather
insights from datasets, but often take a long time to generate.
In this paper, we focus on the problem of 
{\em rapidly generating approximate visualizations
while preserving crucial visual properties of interest to analysts}.
Our primary focus will be on sampling algorithms that preserve the visual property of {\em ordering}; our techniques will
also apply to some other visual properties.
For instance, our algorithms can be used to generate 
an approximate visualization of a bar chart very rapidly, where
the comparisons between any two bars are correct.
We formally show that our sampling algorithms are generally applicable 
and provably optimal in theory, in that they do not take more samples than 
necessary to generate the visualizations with ordering guarantees.
They also work well in practice, 
correctly ordering output groups while
taking orders of magnitude fewer samples and much less time than conventional sampling schemes.}

\end{abstract}

%!TEX root=main.tex

\section{Introduction}\label{sec:intro}

To understand their data, analysts commonly explore their
data\-sets using visualizations, 
often with visual analytics tools 
such as Tableau~\cite{DBLP:conf/sigmod/Hanrahan12} or Spotfire~\cite{spotfire}.
Visual exploration involves
generating a sequence of visualizations, one after the other, 
quickly skimming each one to get a better understanding of
the underlying trends in the datasets.
However, when the datasets are large,
these visualizations often take very long to produce,
creating a significant barrier to interactive analysis.

Our thesis is that on large datasets, we may be able
to quickly produce approximate visualizations of large datasets
preserving visual properties 
crucial for data analysis.
Our visualization schemes will also come with tuning parameters, whereby users can select
the accuracy they desire, choosing less accuracy for more
interactivity and more accuracy for more precise visualizations.

% In this paper, we explore techniques to quickly produce approximate visualizations of large datasets
% that preserve visual properties 
% crucial for data analysis.
% These approximations also provide tuning parameters, whereby users can select
% how much accuracy they desire, choosing less accuracy for more
% interactivity and more accuracy for more precise visualizations.

%Generating these approximate visualizations
%is an acceptable compromise, 
%since data analysts will be able to analyze even larger datasets interactively.
% Our thesis in this paper is that on large datasets,
% we may be able to very quickly generate visualizations that 
% are approximate, i.e., not computed on the entire
% dataset but on a fraction of it, 
% but are still {\em guaranteed to look similar} 
% to the accurate visualization (i.e., the visualization
% computed using the entire dataset). 
% Since analysts are typically more interested in trends and comparisons
% than viewing perfectly accurate visualizations during 
% interactive data analysis, generating
% visualizations that look similar to the accurate visualization 
% is an acceptable compromise, since analysts will be able to 
% analyze even larger datasets interactively.

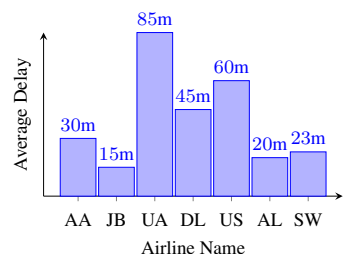
\begin{figure}
\vspace{-5pt}
\centering
\begin{tikzpicture}[thick,scale=0.9, every node/.style={scale=0.9}]
\tikzstyle{every node}=[font=\small]
\begin{axis}[
%    title = Total sales per month,
    ybar,
    bar width=15pt,
    xlabel = {Airline Name},
    ylabel = {Average Delay},
    symbolic x coords = {AA, JB, UA, DL, US, AL, SW},
    ymin=0,
    ytick=\empty,
    xtick = data,
    axis x line=bottom,
    axis y line=left,
    enlarge x limits=0.15,
    xticklabel style={anchor=base,yshift=-\baselineskip},
    nodes near coords={\pgfmathprintnumber\pgfplotspointmeta m}
    ]
    \addplot coordinates {(AA,30) (JB,15) (UA,85) (DL,45) (US,60) (AL, 20) (SW, 23)};
\end{axis}
\end{tikzpicture}
\vspace{-13pt}
\caption{Flight Delays}\label{fig:flt}
\vspace{-23pt}
\end{figure}

We show what we mean by ``preserving visual properties'' 
via an example.
% We will now describe what we mean by ``look similar'' using an example.
Consider the following query
on a database of all flights in the US for the entire year:
\begin{quote}
\vspace{-5pt}
\small
${\tt Q: \ SELECT \ \ NAME, \ \ AVG(DELAY) \ \ FROM \ \ FLT \ \ GROUP \ \ BY \ \ NAME}$
\end{quote}
\vspace{-5pt}
The query asks for the average delays of flights, grouped by 
airline names. 
Figure~\ref{fig:flt} shows a bar chart illustrating an example query result.
In our example, the average delay for AA (American Airlines) 
is 30 minutes, while that for JB (Jet Blue) is
just 15 minutes.
If the ${\tt FLT}$ table is large, the query above 
(and therefore the resulting visualization)
is going to take a very long time to be displayed.
% In fact, recent statistics
% on the number of flights every year in the continental US 
% indicate that the number of tuples in the ${\tt FLT}$
% table could be in the order of several tens of 
% millions [XXX].

In this work, we specifically design {\it sampling
algorithms} that generate  visualizations of queries such as
${\tt Q}$, while sampling only a small fraction of records in the database.
We focus on algorithms that preserve visual properties,
i.e., those that ensure that the visualization appears similar
to the same visualization computed on the entire database. 
% \alkim{This
%   is repetitive. Stating exactly the same thing as 3 paragraphs ago.
%   Also, made me think we were not going to show what ``preserving visual
% properties'' means by example.}
The primary visual property we consider in this paper is the
{\em correct ordering property}: 
ensuring that the groups or bars in a visualization or result set are ordered correctly,
even if the actual value of the group differs from the value that would result if the entire database
were sampled.
% We focus on algorithms that provide {\em visual guarantees}: 
% i.e., those that generate
% visualizations that {\em look similar} to the visualizations
% computed on the entire database. 
% A particularly crucial type of visual guarantee
% that we focus on in this paper
% is a comparison-based guarantee: can we quickly generate
% visualizations where the comparisons between any pairs of groups
% is accurate, i.e., returns the same result as the comparison performed
% on the visualization run on the entire database.
For example, if the delay of JB is smaller than the delay of
AA, then we would like the bar corresponding to JB to
be smaller than the bar corresponding to AA in the output visualization.
As long as the displayed visualizations obey visual properties (such as correct ordering), 
analysts will be able to view trends, gain insights, and make decisions---in our example, 
the analyst can decide which airline should receive
the prize for airline with least delay, or if the analyst sees that
the delay of AL is greater than the delay of SW, they can dig deeper into
AL flights to figure out the cause for higher delay.
Beyond correct ordering, 
our techniques can be applied to other visual properties, including:
\begin{asparaitem}
\item{\bf Accurate Trends}: when generating line charts,
comparisons between neighboring x-axis values must be correctly presented.
\item{\bf Accurate Values}: the values for each group in a bar chart must
be within a certain bound of the values displayed to the analyst.
% \item Accurate Order-of-Magnitudes: if a pair of values are
% very far apart in the accurate visualization, then they must
% not be close in the visualization displayed to the analyst.
% \agp{delete?}
\end{asparaitem}
% Overall, the motivation for the visual guarantees we consider 
% is that analysts are rarely interested in the accurate visualization, 
% and are instead more interested in visualizations that visually look similar.

\begin{figure}
\centering
\begin{tikzpicture}[thick,scale=0.9, every node/.style={scale=0.9}]
\tikzstyle{every node}=[font=\small]
\begin{axis}[
%    title = Total sales per month,
    ybar,
    bar width=15pt,
    xlabel = {Airline Name},
    ylabel = {Average Delay},
    symbolic x coords = {AA, JB, UA, DL, US, AL, SW},
    ymin=0,
    ytick=\empty,
    xtick = data,
    axis x line=bottom,
    axis y line=left,
    enlarge x limits=0.15,
    xticklabel style={anchor=base,yshift=-\baselineskip},
    ]
    \addplot+[error bars/.cd,y dir=both,y explicit]
     coordinates {
    (AA,25) +-  (5, 5) 
    (JB,15) +- (10, 10)
    (UA,85) +- (15, 15)
    (DL,45) +- (10, 10) 
    (US,60) +- (5, 5) 
    (AL, 20) +- (10, 10) 
    (SW, 23) +- (15, 15)};
\end{axis}
\end{tikzpicture}
\vspace{-12pt}
\caption{Flight Delays: Intermediate Processing}\label{fig:flt-bars}
\vspace{-20pt}
\end{figure}
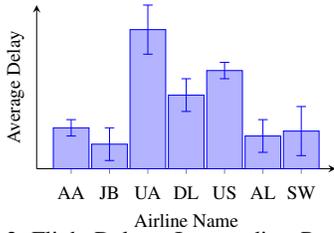

We illustrate the challenges of generating 
accurate visualizations using our flight example.
Here, we assume we have a sampling engine that allows
us to retrieve samples from any airline group at a 
uniform cost per sample (we describe one such sampling
engine we have built in Section~\ref{sec:system}.)
Then, the amount of work done by any visualization generation 
algorithm is proportional
to the number of samples taken in total across all groups.
After performing some work (that is, after doing some sampling), 
let the current state of processing
be depicted as in Figure~\ref{fig:flt-bars}, where
the aggregate for each group is depicted using confidence intervals.
Starting at this point, suppose we wanted to generate a visualization
where the ordering is correct (like in Figure~\ref{fig:flt}).
One option is to use a conventional round-robin stratified sampling
strategy~\cite{casella}, 
\agpneutral{which is the most widely used technique in online approximate 
query processing~\cite{hou89,hou88,lnss93,online-aggregation},}
to take one sample per group in each round,
to generate estimates with shrinking confidence interval bounds. 
This will 
ensure that the eventual aggregate value of each group is roughly correct,
and therefore that the ordering is roughly correct.
We can in fact modify these conventional sampling  
schemes to stop once they are confident that the ordering is guaranteed to be correct. 
However, since conventional sampling is not optimized for 
ensuring that visual properties hold, 
such schemes will end up doing a lot more work than necessary 
(as we will see in the following). 
%We discuss other related work in Section~\ref{sec:related}.

A better strategy would be to focus our attention
on the groups whose confidence intervals continue to overlap with others. 
For instance, for the data depicted in Figure~\ref{fig:flt-bars}, 
we may want to sample more from AA because its
confidence interval overlaps with JB, AL, and SW while
sampling more from UA (even though its confidence interval is large)
is not useful because it gives us no additional information ---
UA is already clearly the airline with the largest delay, even if
the exact value is slightly off.
On the other hand, it is not clear if we should sample more 
from AA or DL, AA has a smaller confidence interval but 
overlaps with more groups, while DL has a larger confidence interval
but overlaps with fewer groups. 
Overall, it is not clear how we may be able to meet our
visual ordering properties while minimizing the samples acquired. 

In this paper, we develop a family of sampling algorithms, based on sound 
probabilistic principles, that:
\begin{asparaenum}
\item are {\em correct}, i.e., they return visualizations 
where the estimated averages are correctly ordered with a probability greater than a user-specified 
threshold, independent of the data distribution,
\item are {\em theoretically optimal}, i.e., no other sampling algorithms can 
take much fewer samples, and
\item are {\em practically efficient}, i.e., they require much fewer samples 
than the size of the datasets to ensure correct visual properties, 
especially on very large datasets. 
In our experiments, our algorithms give us reductions in sampling of
{\em up to 50X} over conventional sampling schemes. 
\end{asparaenum}

\agpneutral{Our focus in this paper is on visualization types that directly correspond
to a SQL aggregation query, e.g., a bar chart, or a histogram; 
these are the most commonly used visualization types in information visualization
applications. 
While we also support generalizations to other visualization types (see Section~\ref{sec:problem-extensions}),
our techniques are not currently applicable to some visualizations, e.g., scatter-plots,
stacked charts, timelines, or treemaps.}

In addition, our algorithms are general enough to retain correctness
and optimality when configured in the following ways:
\begin{asparaenum}
\item {\em Partial Results:} Our algorithms can return partial results 
(that analysts can immediately peruse) improving gradually 
over time.
\item {\em Early Termination:} 
Our algorithms can take advantage of the finite resolution
of visual display interfaces to terminate processing early.
Our algorithms can also terminate early if allowed to make mistakes
on estimating a few groups. 
\item {\em Generalized Settings:}
Our algorithms can be applied to other aggregation functions,
beyond ${\tt AVG}$, as well as other, more complex queries,
and also under more general settings. 
\item {\em Visualization Types:}
Our algorithms can be applied to the generation of 
other visualization types, such as trend-lines or chloropleth maps~\cite{tufte} 
instead of bar graphs.

% \item Can we enforce somewhat stronger considerations than
% comparisons? For instance, if an aggregate for a group
% is actually $k$-times the aggregate for another group 
% in reality, can we ensure that it is within 
% $\alpha k$ and $\frac{k}{\alpha}$ in the visualization 
% that is displayed?

% \item \agp{Not that crucial:} What if, in order to generate
% our visualizations quickly, we are fine with making
% errors on 5\% of the comparisons (in addition to
% 5\% error overall). Firstly, can our 
% algorithms adapt to that case, and how much does
% that buy us?

\end{asparaenum}

\section{Formal Problem Description}\label{sec:setup}

We begin by describing the type of queries and visualizations
that we focus on for the paper. Then, we
describe the formal problem we address.

\subsection{Visualization Setting}
\stitle{Query:} We begin by considering queries such as
our example query in Section~\ref{sec:intro}.
We reproduce the query (more abstractly) here:
\begin{quote}
\vspace{-3pt}
$Q: \ \ {\tt SELECT}  \ \ X, \ \ {\tt AVG}(Y) \ \  {\tt FROM} \ \ R(X, \ \ Y) \ \ {\tt GROUP} \ \ {\tt BY} \ \ X $
\end{quote}
\vspace{-3pt}
This query can be translated to a bar chart visualization
such as the one in Figure~\ref{fig:flt},
where ${\tt AVG}(Y)$ is depicted along the $y-$axis, 
while $X$ is depicted along the $x-$axis.
While we restrict ourselves to queries with a single {\tt GROUP BY} and a {\tt AVG} aggregate,
our query processing algorithms
do apply to a much more general class of queries and visualizations,
including those with other aggregates,
multiple group-bys, and selection or having predicates, as described in Section~\ref{sec:problem-extensions}
(these generalizations still require us to have at least one ${\tt GROUP \ \ BY}$, which restricts us
to aggregate-based visualizations, e.g., histograms, bar-charts, 
and trend-lines.

\stitle{Setting:} 
We assume we have an engine that allows us to efficiently 
retrieve random samples from $R$ corresponding to different values of $X$.
Such an engine is easy to implement, if the relation
$R$ is stored in main memory, and we have a traditional 
(B-tree, hash-based, or otherwise) index on $X$.
We present an approach to implement this engine on disk in Section~\ref{sec:system}.
Our techniques will also apply to the scenario when 
there is no index on $X$ 
\papertext{--- we describe this in the extended technical report~\cite{tr}.}
\techreport{in Section~\ref{sec:ext-no-indexes}.}

% \srm{is this really a necessary assumption?  i think it will make vldb reviewers grumpy.  why not just say that we assume that we have an engine that can efficiently generate samples, which is easy to do in memory and that we present an approach for doing it on disk in section 4.2.3}
% We assume that the relation $R$ is stored in main memory
% with an index on $X$: with increasing memory sizes, 
% it is common to have very large memory machines 
% enabling analysis of ever-increasing volumes of data without having to
% perform disk-reads.
% Note, however, that our techniques will also apply to the scenario
% when there is no index on $X$ (described in Section~\ref{sec:ext-no-indexes}), 
% and when the relation does not fit in main memory 
% (described in Section~\ref{sec:ext-no-memory}). 

\stitle{Notation:} 
We denote the values that the group-by attribute $X$ can take
as $x_1 \ldots x_k$.
We let $n_i$ be the number of tuples in $R$ with $X = x_i$.
For instance, $n_i$ for $X = UA$
will denote the number of flights operated by $UA$ that year.

Let the $i$th group, denoted $S_i$, be
the multiset of the $n_i$ values of $Y$ across all tuples in $R$
where $X = x_i$.
In Figure~\ref{fig:flt}, the 
group corresponding to $UA$ contains the set of delays of
all the flights flown by $UA$ that year.
% By definition, $|S_i| = n_i$.

We denote the {\em true averages} of elements in a group $i$ as $\mu_i$:
Thus, 
$\mu_i = \frac{\sum_{s \in S_i}s}{n_i}$.
The goal for any algorithm processing 
the query $Q$ above is to compute
and display $\mu_i, \forall i \in 1\ldots k$,
such that the estimates for $\mu_i$ are correctly
ordered (defined formally subsequently).

Furthermore, we assume that each value in $S_i$ is bounded
between $[0, c]$. 
For instance, for flights delays, we know that the values
in $S_i$ are within [0, 24 hours], i.e., typical flights are not delayed
beyond 24 hours.
Note however, that our algorithms can still be used when no bound
on $c$ is known, but may not have the desirable 
properties listed in Section~\ref{sec:alg-sample-complexity}.

\subsection{Query Processing}

\stitle{Approach:}
Since we have an index on $X$, we can use 
the index to retrieve a tuple at random with any value of $X = x_i$.
Thus, we can use the index
to get an additional sample of $Y$ at random from any group $S_i$.
Note that if the data is on disk, 
random access through a conventional index can be slow: however, 
we are building a system, called \ntail (also described in Section~\ref{sec:system})
that will address the problem of retrieving samples satisfying arbitrary conditions.

The query processing algorithms that 
we consider take repeated samples from groups $S_i$,
and then eventually output estimates $\nu_1, \ldots,$ $\nu_k$
for true averages $\mu_1, \ldots, \mu_k$.

% \srm{Might be good to note that if the data is on disk, random access through an index can be slow, and to say that
% we are working on this in needletail, or that we aren't solving that problem here.}

\stitle{Correct Ordering Property:}
After retrieving a number of samples,
our algorithm will have some estimate
$\nu_j$ for the value of the actual average $\mu_j$ for each $j$.
When the algorithm terminates and returns the
eventual estimates $\nu_1, \ldots \nu_k$,
we want the following property to hold:
\begin{quote}
\vspace{-3pt}
for all $i, j$ such that $\mu_i > \mu_j$, we have $\nu_i > \nu_j$
\end{quote}
\vspace{-3pt}
We desire that the query processing algorithm always respect
the correct ordering property, but since
we are making decisions probabilistically,
there may be a (typically very small) chance that 
the output will violate the guarantee. 
Thus, we allow the analyst to specify a failure probability
$\delta$ (which we expect to be very close to 0). 
The query processing scheme will then guarantee 
that with probability $1 - \delta$,
the eventual ordering is correct.
We will consider other kinds of guarantees in Section~\ref{sec:problem-extensions}.

\subsection{Characterizing Performance}
We consider three measures for evaluating the performance of 
query processing algorithms:

\stitle{Sample Complexity:}
The cost for any additional sample taken by an algorithm
from any of the groups is the same\footnote{\scriptsize This is certainly
true in the case when $R$ is in memory, but we will describe
why this is true even when $R$ in on disk in Section~\ref{sec:system}.}. 
We denote the total number of samples taken by an algorithm
from group $i$ as $m_i$.
Thus, the total sampling complexity of a query processing
strategy (denoted ${\cal C}$) is the number of samples taken
across groups:
$${\cal C} = \sum_{i\in 1\ldots k} m_i$$

\stitle{Computational Complexity:}
While the total time will be typically dominated by 
the sampling time, we will also analyze the computation 
time of the query processing algorithm, which we
denote ${\cal T}$.

\stitle{Total Wall-Clock Time:} In addition to the two 
complexity measures, we also experimentally 
evaluate the total wall-clock time of 
our query processing algorithms.

\begin{table}
\centering
\scriptsize
\begin{tabular}{|l|l|c|l|c|l|c|l|c|}
\hline
 & \multicolumn{2}{|c|}{Group 1} & \multicolumn{2}{|c|}{Group 2} & \multicolumn{2}{|c|}{Group 3} & \multicolumn{2}{|c|}{Group 4} \\ \hline \hline
1 & [60, 90] & A & [20, 50] & A & [10, 40] & A & [40, 70] & A \\
%10 & [60, 90] & A & [20, 50] & A & [10, 40] & A & [40, 70] & A \\
\hline
\multicolumn{9}{|c|}{\bf \ldots}\\ \hline
20 & [64, 84] & A & [28, 48] & A & [15, 35] & A & [45, 65] & A \\ \hline
21 & [66, 84] & I & [30, 48] & A & [17, 35] & A & [46, 64] & A \\ \hline
\multicolumn{9}{|c|}{\bf \ldots}\\ \hline
57 & [66, 84] & I & [32, 48] & A & [17, 33] & A & [46, 62] & A \\ \hline
58 & [66, 84] & I & [32, 47] & A & [17, 32] & I & [46, 61] & A \\ \hline
\multicolumn{9}{|c|}{\bf  \ldots}\\ \hline
70 & [66, 84] & I & [40, 47] & A & [17, 32] & I & [46, 53] & A \\ \hline
71 & [66, 84] & I & [40, 46] & I & [17, 32] & I & [47, 53] & I \\ \hline
\end{tabular}
\vspace{-5pt}
\caption{Example execution trace: active groups are denoted
using the letter A, while inactive groups are denoted as 
I}\label{tab:exec-trace}
\vspace{-15pt}
\end{table}

\subsection{Formal Problem}
Our goal is to design query processing strategies that preserve 
the right ordering (within the user-specified accuracy bounds)
while minimizing sample complexity:
\begin{problem}[AVG-Order]\label{prob:relative-order}
Given a query $Q$, and parameter values 
$c, \delta$, and an index on $X$, 
design a query processing algorithm 
returning estimates $\nu_1, \ldots, \nu_k$
for $\mu_1, \ldots, \mu_k$
which is as efficient as possible in terms of sample complexity ${\cal C}$,
such that with probability greater than $1 - \delta$,
the ordering of $\nu_1, \ldots, \nu_k$
with respect to $\mu_1, \ldots, \mu_k$
is correct.
\end{problem}

Note that in the problem statement we ignore
computational complexity ${\cal T}$, however,
we do want the computational complexity of our algorithms
 to also be relatively small,
and we will demonstrate that for all algorithms we design, that indeed
is the case.

One particularly important extension we cover right away is the following:
visualization rendering algorithms
are constrained by the number of pixels on the display screen, and
therefore, two groups whose true average values $\mu_i$ are 
{\em very close to each other} cannot be distinguished 
on a visual display screen.
Can we, by relaxing the correct ordering property for groups which are
very close to each other,
get significant improvements in terms of sample and total complexity?
We therefore pose the following problem:
\begin{problem}[AVG-Order-Resolution]\label{prob:relative-order-res}
Given a query $Q$, and values 
$c, \delta$, a minimum resolution $r$, and an index on $X$, 
design a query processing algorithm 
returning estimates $\nu_1, \ldots, \nu_k$
for $\mu_1, \ldots, \mu_k$
which is as efficient as possible in terms of sample complexity ${\cal C}$,
such that with probability greater than $1 - \delta$,
the ordering of $\nu_1, \ldots, \nu_k$
with respect to $\mu_1, \ldots, \mu_k$
is correct, where correctness is now defined as the following:
\begin{quote}
\vspace{-3pt}
for all $i, j, i \neq j$, if $|\mu_i - \mu_j| \leq r$, then ordering
$\nu_i$ before or after $\nu_j$ are both correct, 
while if $|\mu_i - \mu_j| > r$, then
$\nu_i < \nu_j$ if $\mu_i < \mu_j$
and vice versa.
\end{quote}
\end{problem}
\vspace{-3pt}
The problem statement says that if
two true averages, $\mu_i, \mu_j$
satisfy $|\mu_i - \mu_j| \leq r$, then we are no longer
required to order them correctly with respect to each other.

\subsection{Extensions}\label{sec:problem-extensions}
\techreport{In Section~\ref{sec:extensions}}
\papertext{In the technical report~\cite{tr}}
we discuss other
problem variants:
\begin{denselist}
\item Ensuring that weaker properties hold: 
\begin{denselist}
\item {\em Trends and Chloropleths:} When drawing
trend-lines and heat maps (i.e., chloropleths~\cite{tufte}), 
it is more important to ensure  
order is preserved between adjacent groups than between all groups.
\item {\em Top-t Results:} When the number of groups
to be depicted in the visualization is very large, say
greater than 20, it is impossible for users to visually examine all
groups simultaneously. 
Here, the analyst would prefer to view the 
top-$t$ or bottom-$t$ groups in terms of actual averages.
\item {\em Allowing Mistakes:} If the analyst
is fine with a few mistakes being made on a select number of groups (so that 
that the results can be produced faster), this can be
taken into account in our algorithms. 
% \alkim{Is this referring to
% minimum resolution or a different type of extension which allows
% mistakes? We should clarify.}
\end{denselist}
\item Ensuring that stronger properties hold:
\begin{denselist}
\item {\em Values:} We can modify our algorithms to 
ensure that the 
averages $\nu_i$ for each group are close to the actual
averages $\mu_i$, in addition to making sure that
the ordering is correct.
\item {\em Partial Results:} We can modify our algorithms to 
return partial results as an when they are computed.
This is especially important when the visualization takes 
a long time to be computed, so that
the analyst to start perusing the visualization as soon
as possible.
\end{denselist}
\item Tackling other queries or settings:
\begin{denselist}
\item {\em Other Aggregations:} We can generalize our techniques
for aggregation functions beyond ${\tt AVG}$,
including ${\tt SUM}$ and ${\tt COUNT}$.
\item {\em Selection Predicates:} Our techniques apply
equally well when we have WHERE or HAVING predicates 
in our query.
\item {\em Multiple Group Bys or Aggregations:} We can generalize
our techniques to handle the case where we are visualizing multiple
aggregates simultaneously, and when we are grouping by multiple 
attributes at the same time (in a three dimensional visualization
or a two dimensional visualization with a cross-product on the x-axis).
\item {\em No indexes:} Our techniques also apply to the scenario when
we have no indexes.
\end{denselist}
\end{denselist}

%!TEX root=main.tex

% \begin{algorithm}[h!]
% \KwData{$S_1,\ldots,S_k,\delta$}
% %\KWResult{$\nu_1,\ldots,\nu_k$}
% Initialize $m \gets 10$\;
% Draw $m$ samples from each of $S_1,\ldots,S_k$ to provide initial estimates $\nu_1,\ldots,\nu_k$\;  
% Initialize $A = \{1, \ldots, k\}$\;
% \While{$A \neq \emptyset$}{
%   $m \gets m+1$\;
%   $\varepsilon = c \sqrt{\frac{\log(4k|A|/\delta)}{2m}}$\;
%   \For{each $i \in A$}{
%     Draw a sample $x$ from $S_i$\;
%     $\nu_i \gets \frac{m-1}{m} \nu_i + \frac1{m}x$\;
%     }
%   \For{each $i \in A$}{
%     \If{$[\nu_i - \varepsilon, \nu_i + \varepsilon] \cap 
%           \big(\bigcup_{j \in A \setminus \{i\}} [\nu_j - \varepsilon, \nu_j + \varepsilon]\big) = \emptyset$}{$A \gets A \setminus \{i\}$}
%     } 
% }
% Return $\nu_1,\ldots,\nu_k$\;
% \caption{\ifocus}\label{alg:pseudocode}
% \end{algorithm}

\begin{algorithm}[h!]
\caption{\ifocus}\label{alg:pseudocode}
\KwData{$S_1,\ldots,S_k,\delta$}
Initialize $m \gets 1$\;
Draw $m$ samples from each of $S_1,\ldots,S_k$ to provide initial estimates $\nu_1,\ldots,\nu_k$\;  
Initialize $A = \{1, \ldots, k\}$\;
\While{$A \neq \emptyset$}{
  $m \gets m+1$\;
  $\varepsilon = c \, \sqrt{(1-\frac{m/\kappa-1}{\max_{i \in A} n_i})
  \frac{2\log\log_\kappa(m) + \log(\pi^2k/3\delta)}{2m/\kappa}}$ /$\ast$Update Confidence Interval Size$^\dagger$ $\ast$/\; 
  \For{each $i \in A$}{
    Draw a sample $x$ from $S_i$\;
    $\nu_i \gets \frac{m-1}{m} \nu_i + \frac1{m}x$\;
    }
  \For{each $i \in A$}{
    \If{$[\nu_i - \varepsilon, \nu_i + \varepsilon] \cap 
          \big(\bigcup_{j \in A \setminus \{i\}} [\nu_j - \varepsilon, \nu_j + \varepsilon]\big) = \emptyset$}{$A \gets A \setminus \{i\}$}
    } 
}
Return $\nu_1,\ldots,\nu_k$\;
\end{algorithm}

\blfootnote{
  $^\dagger${\scriptsize We are free to set $\kappa$ to any number greater than $1$;
  in our experiments, we set $\kappa = 1$. Since this would render
  $\log_{\kappa}$ infinite, for that term, we use $\log_{e}$. We found
  that setting $\kappa$ equal to a small value close to $1$ (e.g.,
  $1.01$) gives very similar results on both accuracy and latency since
  the term that dominates the sum in the numerator is not the $\log
  \log_{\kappa} m$.}
}

\vspace{-15pt}
\section{The algorithm and its analysis}\label{sec:solution}

In this section, we describe our solution to Problem~\ref{prob:relative-order}.
We start by introducing the new
\ifocus\ algorithm in
Section~\ref{sec:alg-description}.
We will analyze its sample complexity 
and demonstrate its correctness in Section~\ref{sec:alg-sample-complexity}.
We will then analyze its computational complexity
in Section~\ref{sec:alg-comp-complexity}.
Finally, we will demonstrate that the \ifocus\ algorithm
is essentially optimal, i.e., no other algorithm
can give us a sample complexity much smaller than \ifocus,
in Section~\ref{sec:alg-lower-bound}.

\subsection{The Basic {\large \ifocus} Algorithm}\label{sec:alg-description}
The  \ifocus\ algorithm is shown in Algorithm~\ref{alg:pseudocode}.
We describe the pseudocode and illustrate the execution on an example below. 

At a high level, the algorithm works as follows.  For each group, it maintains a confidence interval 
(described in more detail below)
within which the algorithm believes the true average of each group lies.
The algorithm then proceeds in rounds.
The algorithm starts off with one sample per group
to generate initial confidence intervals for the true averages $\mu_1, \ldots, \mu_k$.
We refer to the groups whose confidence intervals overlap with other groups
as {\em active groups}.
Then, in each round, for all the groups whose confidence intervals still
overlap with confidence intervals of other groups, i.e., 
all the active groups, 
a single additional sample is taken.
% We keep taking samples from the active groups,
% until there are no more active groups.
We terminate when there are no remaining active groups
and then return the estimated averages $\nu_1, \ldots, \nu_k$.
We now describe an illustration of the algorithm on an example.
%\srm{what other aggregates does it generalize for?  SUM?  COUNT? STD DEV?}

\begin{table}[t!]
\scriptsize
\centering
\begin{tabular}{|l||p{5cm}|}
\hline
$k$ & Number of groups. \\ \hline
$n_1,\ldots,n_k$ & Number of elements in each group. \\ \hline
$S_1,\ldots,S_k$ & The groups themselves. $S_i$ is a set of $n_i$ elements from $[0,1]$. \\ \hline
$\mu_1,\ldots,\mu_k$ &  Averages of the elements in each group. $\mu_i = \E_{x \in S_i}[x]$. \\ \hline
$\tau_{i,j}$ & Distance between averages $\mu_i$ and $\mu_j$. $\tau_{i,j} = |\mu_i - \mu_j|$. \\ \hline
$\eta_i$ & Minimal distance between $\mu_i$ and the other averages. $\eta_i = \min_{j \neq i} \tau_{i,j}$. \\ \hline
$r$ &  Minimal resolution, $0 \le r \le 1$. \\ \hline
$\eta_i^{(r)}$ & Thresholded minimal distance; $\eta_i^{(r)} = \max\{\eta_i, r\}$. \\ \hline
\end{tabular}
\vspace{-7pt}
\caption{Table of Notation}\label{tab:notation}
\vspace{-15pt}
\end{table}

\begin{example}
An example of how the algorithm works
is given in Table~\ref{tab:exec-trace}.
Here, there are four groups, i.e., $k = 4$.
Each row in the table corresponds to one phase of sampling.
The first column refers to the total number of samples 
that have been taken so far for each
of the active groups (we call this the number of the round).
The algorithm starts by taking one sample per group 
to generate initial confidence intervals:
these are displayed in the first row.

At the end of the first round, all four groups are active since for every confidence interval, there
is some other confidence interval with which it overlaps.
For instance, for group 1, whose confidence interval is [60, 90], this
confidence interval overlaps with the confidence interval of group 4;
therefore group $1$ is active.

We ``fast-forward'' to  round 20, where once again all groups are still active.
Then, on round 21, after an additional sample, the confidence interval of group 1 shrinks to [66, 84],
which no longer overlaps with any other confidence interval.
Therefore, group 1 is no longer active, and we stop sampling from group 1.
We fast-forward again to round 58, where after taking a sample, group 3's confidence interval
no longer overlaps with any other group's confidence interval, so we can stop sampling it too.
Finally, at round 71, none of the four confidence intervals overlaps with any other.
Thus, the total cost of the algorithm (i.e., the number of samples) is
$$ {\cal C}  = 21 \times 4 + (58 - 21) \times 3 + (71 - 58) \times 2 $$
The expression $21 \times 4$ comes from the 21 rounds when all four groups are active,
$(58 - 21) \times 3$ comes from the rounds from 22 to 58, when only three groups are active,
and so on.
\end{example}

The pseudocode for the algorithm is shown in Algorithm~\ref{alg:pseudocode}; $m$ refers to the round.
We start at the first round (i.e., $m = 1$)
drawing one sample from each of $S_1, \ldots, S_k$
to get initial estimates of $\nu_1, \ldots, \nu_k$.
Initially, the set of active groups, $A$, contains all groups from $1$ to $k$.
As long as there are active groups,
in each round, we take an additional sample for all the groups in $A$,
and update the corresponding $\nu_i$.
Based on the number of samples drawn per active group, we update $\varepsilon$, i.e., 
the half-width of the confidence interval.
Here, the confidence interval $[\nu_i - \varepsilon, \nu_i + \varepsilon]$ refers
to the $1-\delta$ confidence interval on taking $m$ samples, i.e., 
having taken $m$ samples, the probability that the true average $\mu_i$
is within $[\nu_i - \varepsilon, \nu_i + \varepsilon]$ is greater than $1 - \delta$.
As we show below, the confidence intervals are derived using 
a variation of Hoeffding's inequality.

\stitle{Discussion:}
We note several features of the algorithm:

\begin{denselist}
\item 
As we will see, the sampling complexity of \ifocus\ does not depend on
the number of elements in each group, and simply depends on where
the true averages of each group are located relative to each other.
We will show this formally in Section~\ref{sec:alg-sample-complexity}.
%\alkim{This isn't 100\% true; when calculating $\varepsilon$, there is a
%dependence on $n_i$}

\techreport{
\item The algorithm has similar
guarantees and properties when the sampling per group is done with 
as against without replacement. 
We will discuss these differences 
%in Section~\ref{sec:alg-sample-complexity}.
in Section~\ref{sec:discussion}.
}

% Since sampling with replacement is easier to implement (because we do not
% need to keep track of the values seen so far), 
% we will focus on sampling with replacement for the rest of the paper.

\item There is a corner case that needs to be treated carefully: 
there is a small chance that a group that was not active 
suddenly becomes active because the average $\nu_i$ 
of some other group moves excessively 
due to the addition of a very large (or very small) element. 
We have two alternatives at this point
\begin{denselist}
\item a) ignore the newly activated group; i.e., groups can never be
  added back to the set of active groups
\item b) allow inactive groups to become active.
\end{denselist}
It turns out the properties we prove for the algorithm in terms of optimality
of sample complexity (see Section~\ref{sec:alg-sample-complexity})
hold if we do a).
If we do b), the properties of optimality no longer hold.

% That said, in practice, we prefer doing b) because
% the case where an inactive set becomes active is likely 
% to be a case where the algorithm may end up making an error. 
% \srm{I am confused by the previous statement. How do we know this is true?  Sounds like we are saying we don't care if it makes
% an error.  As long as we are appealing to intuition, I would rather make a claim based on real data -- i.e., say how frequently this occurs in our actual data and claim that suggests it
% doesn't really matter in practice so we do b) because it is easier.}
% Thus, in our implementation, we do b) instead of a), even though a) has
% provably optimal sample complexity.

\end{denselist}

\subsection{Proof of Correctness}
We now prove that \ifocus\ obeys the ordering property
with probability greater than $1-\delta$. 
Our proof involves three steps:
\begin{denselist}
\item {\bf Step 1:} The algorithm \ifocus\ 
obeys the correct ordering property, as long as 
the confidence intervals
of each active group contain the actual average,
during every round.
\item {\bf Step 2:} The confidence intervals
of {\em any given} active group contains
the actual average of that group with
probability greater than $(1 - \delta/k)$ 
at every round, as long as
$\varepsilon$ is set according to Line
6 in Algorithm~\ref{alg:pseudocode}.
\item {\bf Step 3:} The confidence intervals 
of {\em all} active groups contains actual averages
for the groups with probability
greater than $(1 - \delta)$ 
at every round,
when $\varepsilon$ is set
as per Line 6 in Algorithm~\ref{alg:pseudocode}.
\end{denselist}
Combining the three steps together give us the desired result.

\stitle{Step 1:} To complete this step, 
we need a bit more notation. 
For every $m > 1$, let $A_m$, $\varepsilon_m$, and 
$\nu_{1,m},\ldots,\nu_{k,m}$ denote the values of 
$A,\varepsilon,\nu_1,\ldots,\nu_k$ at step 10
in the algorithm for the iteration of the loop corresponding to $m$.
Also, for $i = 1,\ldots,k$, recall that $m_i$ is the number
of samples required to estimate $\nu_i$;  
equivalently, it will denote the value of 
$m$ when $i$ is removed from $A$.
We define $m_{max}$ to be the largest $m_i$.
\begin{lemma}
\label{lem:correct}
If for every $m \in 1\ldots m_{max}$ and 
every $j \in A_{m}$, 
we have $|\nu_{j,m} - \mu_j| \le \varepsilon_{m}$,
then the estimates $\nu_1,\ldots,\nu_k$ 
returned by the algorithm have the same order as 
$\mu_1,\ldots,\mu_k$, i.e., the algorithm satisfies the correct ordering property.
\end{lemma}
That is, as long as all the estimates for the 
active groups are close enough to their true average, 
that is sufficient to ensure overall
correct ordering.

\begin{proof}
Fix any $i \neq j \in \{1, \ldots, k\}$. 
We will show that $\nu_i > \nu_j$ iff
$\mu_i > \mu_j$. 
Applying this to all $i, j$ gives us
the desired result.

Assume without loss of generality 
(by relabeling $i$ and $j$, if needed) 
that $m_i \leq m_j$.
Since $m_i \leq m_j$, $j$ is removed
from the active groups at a later stage than $i$.
At $m_i$, we have that the confidence
interval for group $i$ no longer overlaps 
with other confidence intervals (otherwise
$i$ would not be removed from the set of active
groups).
Thus, the intervals $[\nu_{i,m_i} - \varepsilon_{m_i}, 
\nu_{i,m_i} + \varepsilon_{m_i}]$ and 
$[\nu_{j,m_i} - \varepsilon_{m_i}, \nu_{j,m_i} + \varepsilon_{m_i}]$
are disjoint.
Consider the case when $\mu_i < \mu_j$.
Then, we have: 
\begin{align}
\mu_i \leq \nu_{i,m_i} + \varepsilon_{m_i} &  < \nu_{j,m_i} - \varepsilon_{m_i}  \le \mu_j \label{eq:general-containment} \\
\implies \nu_{i, m_i} & < \mu_j - \varepsilon_{m_i} \label{eq:containment}
\end{align}
The first and last inequality 
holds because $\mu_i$ and
$\mu_j$ are within
the confidence interval around $\nu_i$ and
$\nu_j$ respectively at round $m_i$.
The second inequality holds because
the intervals are disjoint. 
(To see this, notice that if the
inequality was reversed, the intervals would
no longer be disjoint.)
Then, we have: 
\begin{align}
\nu_j = \nu_{j,m_j} \ge \mu_j - \varepsilon_{m_j} \ge \mu_j - \varepsilon_{m_i} > \nu_{i,m_i} = \nu_i.\label{eq:ordering-property}
\end{align}
The first equality holds because
group $j$ exits the set of active groups at
$m_j$;
the second inequality holds because
the confidence interval at $j$
contains $\mu_j$;
the third inequality holds because
$\varepsilon_j \leq \varepsilon_i$
(since confidence intervals shrink
as the rounds proceed);
the next inequality holds because 
of Equation~\ref{eq:containment};
while the last equality holds
because group $i$ exits
the set of active groups at $m_i$.
Therefore, we have $\nu_i < \nu_j$, as desired.
The case where $\mu_i > \mu_j$ 
is essentially identical: in this case Equation~\ref{eq:general-containment} 
is of the form:
$$
\mu_i \ge \nu_{i,m_i} - \varepsilon_{m_i} > \nu_{j,m_i} + \varepsilon_{m_i} \ge \mu_j
$$
and Equation~\ref{eq:ordering-property} is of the form:
$$
\nu_j = \nu_{j,m_j} \le \mu_j + \varepsilon_{m_j} \le \mu_j + \varepsilon_{m_i} < \nu_{i,m_i} = \nu_i.
$$
so that we now have $\nu_i > \nu_j$, once again
as desired.
\end{proof}

\stitle{Step 2:} In this step, our goal is to prove
that the confidence interval of any group contains
the actual average with probability greater than $(1 - \delta/k)$
on following Algorithm~\ref{alg:pseudocode}.

For this proof, we use a specialized concentration inequality 
that is derived from Hoeffding's classical inequality~\cite{all-of-statistics}.
Hoeffding~\cite{hoeffding1963probability} showed that 
his inequality can be applied to
this setting to bound the deviation of the average of  
random numbers sampled from a set 
from the true average of the set.
Serfling~\cite{serfling1974probability} refined 
the previous result to give tighter bounds 
as the number of random numbers sampled
approaches the size of the set.

\begin{lemma}[Hoeffding--Serfling inequality~\cite{serfling1974probability}]
\label{lem:h-s}
Let $\calY = y_1,$ $\ldots,$ $y_N$ be a set of $N$ 
values in $[0,1]$ with average value $\frac1N \sum_{i=1}^N y_i = \mu$.
Let $Y_1,\ldots,Y_N$ be a 
sequence of random variables drawn from $\calY$ without
replacement. 
For every $1 \le k < N$ and $\varepsilon > 0$,
$$
\Pr\left[ \max_{k \le m \le N-1} \left|\frac{\sum_{i=1}^m Y_i}{m} - \mu\right| \ge \varepsilon \right]
\le 2 \exp\left( - \frac{2k\varepsilon^2}{1-\frac{k-1}N} \right).
$$
\end{lemma}
We use the above inequality to get tight bounds for the 
value of $\sum_{i=1}^m Y_i/m$ for all $1 \le m \le N$, with 
probability $\delta$. 
We discuss next how to apply the theorem to
complete Step 2 of our proof.
\begin{theorem}
\label{thm:hs}
Let $\calY = y_1,$ $\ldots,$ $y_N$ be a set of $N$ 
values in $[0,1]$ with average value
$\frac1N \sum_{i=1}^N y_i = \mu$.
Let $Y_1,\ldots,Y_N$ be a 
sequence of random variables drawn from $\calY$ without
replacement.
Fix any $\delta > 0$ and $\kappa > 1$. For $1 \le m \le N-1$, define
$$
\varepsilon_m = \sqrt{\frac{(1-\frac{m/\kappa-1}N)(2\log \log_\kappa (m) + \log(\pi^2/3\delta))}{2m/\kappa}}.
$$
$$
\text{Then:} \ \Pr\left[ \exists m, 1 \le m \le N : 
  \left|\frac{\sum_{i=1}^m Y_i}{m} - \mu\right| > \varepsilon_m \right] 
\le \delta.
$$
\end{theorem}

\begin{proof}
We have:
{\small
\begin{align*}
& \Pr\left[ \exists m, 1 \le m \le N : 
  \left|\frac{\sum_{i=1}^m Y_i}{m} - \mu\right| > \varepsilon_m \right] \\
&\le \sum_{r \ge 1} \Pr\left[ \exists m, \kappa^{r-1} \le m \le \kappa^r : \left|\frac{\sum_{i=1}^m Y_i}{m} - \mu\right| > \varepsilon_m \right] \\
&\le \sum_{r \ge 1} \Pr\left[ \exists m, \kappa^{r-1} \le m \le \kappa^r : \left|\frac{\sum_{i=1}^m Y_i}{m} - \mu\right| > \varepsilon_{\kappa^{r}} \right] \\
&\le \sum_{r \ge 1} \Pr\left[ \max_{\kappa^{r-1} \le m \le N-1} \left|\frac{\sum_{i=1}^m Y_i}{m} - \mu\right| > \varepsilon_{\kappa^r} \right].
\end{align*}
}
The first inequality holds by the union bound~\cite{all-of-statistics}
\techreport{(i.e., the probability that a union of events occurs
is bounded above by sum of the probabilities that each occurs)}.
The second inequality holds because $\varepsilon_m$ 
only decreases as $m$ increases.
The third inequality  
holds because the condition that 
any of the sums on the left-hand side is greater than 
$\varepsilon_{\kappa^r}$ occurs when 
the maximum is greater than $\varepsilon_{\kappa^r}$.
%In general, as 
%$\kappa$ decreases, $\epsilon_{\kappa^r}$ also decreases, and we are
%able to achieve smaller sample complexities. For our experiments, we
%take the limit and set $\kappa = 1$. For $\log_{\kappa}$, we
%we make an approximation using the natural log and show in our
%experiments that this approximation does not deteriorate our accuracy.

By the Hoeffding--Serfling inequality (i.e., Lemma~\ref{lem:h-s}), 
$$
\Pr\left[ \max_{\kappa^{r-1} \le m \le N-1} \left|\frac{\sum_{i=1}^m Y_i}{m} - \mu\right| > \varepsilon_{\kappa^r} \right]
\le \frac{6\delta}{\pi^2 r^2}.
$$
The theorem the follows from
the identity $\sum_{r \ge 1} \frac1{r^2} = \pi^2/6$.
\end{proof}
Now, when we apply Theorem~\ref{thm:hs} to any group $i$ in
Algorithm~\ref{alg:pseudocode}, with $\varepsilon_m$ set as described
in Line 6 in the algorithm,
$N$ set to $n_i$, 
$Y_i$ being equal to the $i$th
sample from the group (taken without replacement), 
and $\delta$ set to $\delta/k$, 
we have the following corollary.
\begin{corollary}\label{cor:per-group-guarantee}
For any group $i$, across all rounds of Algorithm~\ref{alg:pseudocode},
we have:
$
\Pr\left[ \exists m, 1 \le m \le m_i : 
  \left|\nu_{i, m} - \mu\right| > \varepsilon_m \right] 
\le \delta/k.
$
\end{corollary}
\stitle{Step 3:} On applying the union bound~\cite{all-of-statistics}
to Corollary~\ref{cor:per-group-guarantee},
we get the following result:
\begin{corollary}\label{cor:all-group-guarantee}
Across all groups and rounds of Algorithm~\ref{alg:pseudocode}:
$
\Pr\left[ \exists i, m,  1\leq i \leq k, 1 \le m \le m_i : 
  \left|\nu_{i, m} - \mu\right| > \varepsilon_m \right] 
\le \delta.
$
\end{corollary}
This result, when combined with Lemma~\ref{lem:correct},
allows us to infer the following theorem:
\vspace{-7pt}
\begin{framed}
\vspace{-15pt}
\begin{theorem}[Correct Ordering]
The eventual estimates $\nu_1, \ldots, \nu_k$ 
returned by Algorithm~\ref{alg:pseudocode} have
the same order as $\mu_1, \ldots, \mu_k$
with probability greater than $1-\delta$.
\end{theorem}
\vspace{-15pt}
\end{framed}
\vspace{-7pt}

\subsection{Sample Complexity of {\large \ifocus}}\label{sec:alg-sample-complexity}

To state and prove the theorem about the sample complexity of \ifocus,
we introduce some additional notation which 
allows us to describe the ``hardness'' of a particular input instance.
(Table~\ref{tab:notation} describes all the symbols used in the paper.)
We define $\eta_i$ to be the minimum distance between $\mu_i$
and the next closest average,
i.e., $\eta_i = \min_{j \neq i}{|\mu_i - \mu_j|}$.
The smaller $\eta_i$ is, the more effort we  need 
to put in to ensure that the confidence interval estimates for $\mu_i$
are are small enough compared to $\eta_i$.

In this section, we prove the following theorem:
\vspace{-7pt}
\begin{framed}
\vspace{-15pt}
\begin{theorem}[Sample Complexity]
\label{thm:analysis}
With probability at least $1-\delta$,  \ifocus\ 
outputs estimates $\nu_1,\ldots,\nu_k$ that satisfy the correct ordering property 
and, furthermore, draws 
\begin{align}\label{eq:samp-complexity}
O\left( c^2 \sum_{i=1}^k \frac{\log(\frac{k}{\delta}) + \log \log(\frac{1}{\eta_i})}{\eta_i^{\,2}}\right) \ \text{samples in total.}
\end{align}
\end{theorem}
\vspace{-15pt}
\end{framed}
\vspace{-7pt}
The theorem states that \ifocus\ obeys
the correct ordering property while drawing a number of samples
from groups proportional to the sum of the inverse of the squares of the $\eta_i$:
that is, the smaller the $\eta_i$, the larger the amount of sampling
we need to do (with quadratic scaling).

% We prove the theorem using Hoeffding's inequality.

% \begin{theorem}[Hoeffding's inequality]
% Let $X_1,$ $\ldots,$ $X_m$ be i.i.d.~random variables bounded by $[0,1]$ and with mean $\E X_1 = \mu$.
% Then for every $t > 0$,
% $$
% \Pr\left[ \left|\frac{X_1+ \cdots + X_m}{m} - \mu\right| > t \right] \le 2 e^{- 2 t^2 m}.
% $$
% \end{theorem}
% The first step of the analysis of the algorithm applies Hoeffding's inequality to show that 
% for any fixed $m$ and any $i \in A$, the interval $[\nu_i - \varepsilon, \nu_i + \varepsilon]$
% contains $\mu_i$ with high probability.

% \begin{corollary}
% \label{cor:hoeffding}
% For any $m \ge 10$, any $i \in A$, and any $\gamma > 0$, 
% $$
% \Pr\left[ |\nu_i - \mu_i| \le c \sqrt{\frac{\log(2/\gamma)}{2m}} \right] \ge 1 - \gamma.
% $$
% \end{corollary}

% \begin{proof}
% This is just a restatement of Hoeffding's inequality since $\nu_i$ is the empirical
% average of $m$ i.i.d.~random variables bounded by $[0,c]$ and with mean $\mu_i$.
% \end{proof}

The next lemma gives us an upper bound on how large $m_i$ can be 
in terms of the $\eta_i$, for each $i$: this allows us to establish an upper bound
on the sample complexity of the algorithm.

\begin{lemma}
\label{lem:samples}
Fix $i \in 1 \ldots k$. Define $m_i^*$ to be the minimal value of $m \ge 1$ for which
$\varepsilon_m < \eta_i/4$. In the running of the algorithm, 
if for every $j \in A_{m_i^*}$, 
we have $|\nu_{j,m_i^*} - \mu_j| \le \varepsilon_{m_i^*}$, then $m_i \le m_i^*$.
\end{lemma}
Intuitively, the lemma allows us to establish that $m_i < m_i^*$, the latter
of which (as we show subsequently) is dependent on $\eta_i$.
\begin{proof}
If $i \notin A_{m_i^*}$, then the conclusion of the lemma trivially holds,
because $m_i < m_i^*$. 
Consider now the case where $i \in A_{m_i^*}$. 
We now prove that $m_i = m_i^*$.
Note that $m_i = m_i^*$ if and only if the interval
$[\nu_{i,m_i^*} - \varepsilon_{m_i^*}, \nu_{i,m_i^*} + \varepsilon_{m_i^*}]$ is disjoint
from the union of intervals $\bigcup_{j \in A_{m_i^*} \setminus \{i\}}
[\nu_{j,m_i^*} - \varepsilon_{m_i^*}, \nu_{j,m_i^*} + \varepsilon_{m_i^*}]$.

We focus first on all $j$ where $\mu_j < \mu_i$.
By the definition of $\eta_i$, every $j \in A_{m_i^*}$ for which
$\mu_j < \mu_i$ satisfies the stronger inequality $\mu_j \le \mu_i - \eta_i$.
By the conditions of the lemma (i.e., that
confidence intervals always contain the true average),  
we have that $\mu_j \ge \nu_{j,m_i^*} - \varepsilon_{m_i^*}$
and that $\mu_i \le \nu_{i,m_i^*} + \varepsilon_{m_i^*}$. 
So we have:
{\scriptsize
\vspace{-3pt}
\begin{align*}
\nu_{j,m_i^*} + \varepsilon_{m_i^*}  \le   \mu_j + 2\varepsilon_{m_i^*}   <  \mu_j + \frac{\eta_i}{2}   \le  \mu_i - \frac{\eta_i}{2} < \mu_i - 2\varepsilon_{m_i^*}  \le   \nu_{i,m_i^*} - \varepsilon_{m_i^*}
\end{align*}
}
\vspace{-3pt}
\begin{denselist}
\item The first and last inequalities follow the fact that 
the confidence interval for $\nu_j$ always contains $\mu_j$,
i.e., $\mu_j \geq \nu_{j, m_i^*} - \varepsilon_{m_i^*}$;
\item the second and fourth follow from the fact that $\varepsilon_{m_i^*} < \eta_i/4$; 
\item and the third follows from the fact that $\mu_j \leq \mu_i - \eta_i$.
\end{denselist}
Thus, the intervals $[\nu_{i,m_i^*} - \varepsilon_{m_i^*}, \nu_{i,m_i^*} + \varepsilon_{m_i^*}]$ and
$[\nu_{j,m_i^*} - \varepsilon_{m_i^*}, \nu_{j,m_i^*} + \varepsilon_{m_i^*}]$ are disjoint.
Similarly, for all $j \in A_{m_i^*}$ that satisfies $\mu_j > \mu_i$, 
we observe that the interval 
$[\nu_{i,m_i^*} - \varepsilon_{m_i^*}, \nu_{i,m_i^*} + \varepsilon_{m_i^*}]$
is also disjoint from 
$[\nu_{j,m_i^*} - \varepsilon_{m_i^*}, \nu_{j,m_i^*} + \varepsilon_{m_i^*}]$.
\end{proof}

We are now ready to complete the analysis of the algorithm.

\begin{proof}[of Theorem~\ref{thm:analysis}]
First, we note that for $i =1,\ldots,k$, the value $m_i^*$ is bounded above by
$$
m_i^* = O\left( c^2 \frac{\log \log \frac1{\eta_i^{\,2}} + \log \frac{k}{\delta}}{\eta_i^{\,2}} \right).
$$
(To verify this fact, note that when $m = \frac{8c^2}{\eta_i^{\,2}}(\log \frac{\pi^2 k}{3\delta} + \log \log \frac{8}{\eta_i{\,2}} + 1)$, 
then the corresponding value of $\varepsilon$ satisfies
$\varepsilon_m < \frac{\eta_i}{4}$.)

By Corollary~\ref{cor:all-group-guarantee}, with probability at least $1-\delta$,
for every $i \in 1,\ldots,k$, every $m \ge 1$, and every $j \in A_{m}$, 
we have $|\nu_{j,m} - \mu_j| \le \varepsilon_{m}$.
Therefore, by Lemma~\ref{lem:correct} the estimates $\nu_1,\ldots,\nu_k$ returned
by the algorithm satisfy the correct ordering property.
Furthermore, by Lemma~\ref{lem:samples}, 
the total number of samples drawn from the $i$th 
group by the algorithm is bounded above by $m_i^*$ and 
the total number of samples requested by the algorithm is bounded above by
\[
\sum_{i=1}^k m_i^* = O\left( c^2 \sum_{i=1}^k \frac{\log(\frac{k}{\delta}) + \log \log(\frac1{\eta_i})}{\eta_i^{\,2}} \right). 
\]
We have the desired result.
\end{proof}

\subsection{Computational Complexity}\label{sec:alg-comp-complexity}
The computational complexity of the algorithm
is dominated by the check used to determine if a group is still active.
This check can be done in 
$O(\log |A|)$ time per round if we maintain a binary search tree --- 
leading to $O(k \log k)$ time per round across
all active groups.
However, in practice, $k$ will be small (typically less than 100); 
and therefore, taking an additional sample from a group will dominate the
cost of checking if groups are still active.

Then, the number of rounds is the largest value 
that $m$ will take in Algorithm~\ref{alg:pseudocode}.
This is in fact:
$$(\log{\frac{k}{\delta}} + \log{\frac{1}{\eta}}) \frac{c^2}{\eta^2},$$
where $\eta = \min_i{\eta_i}$.
Therefore, we have the following theorem:
\vspace{-7pt}
\begin{framed}
\vspace{-15pt}
\begin{theorem}
The computational complexity of the \ifocus\ algorithm is:
$O(k \log (k) (\log{\frac{k}{\delta}} + \log{\frac{1}{\eta}}) \frac{c^2}{\eta^2})$.
\end{theorem}
\vspace{-15pt}
\end{framed}
\vspace{-7pt}

\subsection{Lower bounds on Sample Complexity}\label{sec:alg-lower-bound}

We now show that the sample complexity of 
\ifocus\ is optimal as compared to any algorithm
for Problem~\ref{prob:relative-order}, up to a small
additive factor, and constant multiplicative factors.

\vspace{-7pt}
\begin{framed}
\vspace{-15pt}
\begin{theorem}[Lower Bound]
\label{thm:lb}
Any algorithm that satisfies the correct ordering condition with probability
at least $1 - \delta$ must make at least $\Omega(\log(\frac{k}{\delta}) \sum_{i=1}^k
\frac{c^2}{\eta_i^{\,2}} )$ queries.
\end{theorem}
\vspace{-15pt}
\end{framed}
\vspace{-7pt}
Comparing the expression above to Equation~\ref{eq:samp-complexity},
the only difference is a small additive term: 
$\frac{c^2}{\eta_i^2} \log \log (\frac1{\eta_i})$,
which we expect to be much smaller than $\frac{c^2}{\eta_i^2} \log (\frac{k}{\delta})$.
Note that even when $\frac1{\eta_i}$ is $10^9$
(a highly unrealistic scenario), we have that $\log \log \frac1{\eta_i} < 5$,
whereas $\log \frac{k}{\delta}$ is greater
than $5$ for most practical cases (e.g., when $k = 10, \delta = 0.05$).

The starting point for our proof of this theorem is a lower bound for sampling
due to Canetti, Even, and Goldreich~\cite{canetti}.

\begin{theorem}[Canetti--Even--Goldreich~\cite{canetti}]
Let $\varepsilon \le \frac18$ and $\delta \le \frac16$. Any algorithm that
estimates $\mu_i$ within error $\pm \varepsilon$ with confidence $1-\delta$ must
sample at least $\frac{1}{8\varepsilon^2} \ln( \frac1{16e\sqrt{\pi} \delta} )$
elements from $S_i$ in expectation.
\end{theorem}
In fact, the proof of this theorem yields a slightly stronger result: even
if we are promised that $\mu_i \in \{\frac12 - \varepsilon, \frac12 + \varepsilon\}$, 
the same number of samples is required to distinguish between the two cases.
% And I believe that essentially the same bound also holds if we are promised
% that $\mu_i \in \{c - \varepsilon, c + \varepsilon\}$ for any constant value $c$
% in $(0,1)$. (The proof of this extension needs to be written down.)

\papertext{The proof of Theorem~\ref{thm:lb} is omitted due to lack of space, and can be
found in the extended technical report~\cite{tr}.}
\techreport{
\begin{proof}[of Theorem~\ref{thm:lb}]
We prove the theorem using $c = 1$ (where $c$ is the upperbound
for any individual element in a group). 
It is easy to modify the proof for when $c \neq 1$.

Fix some distance $\tau < 1/20k$. Let $S_1,\ldots,S_{k/2}$ be sets of elements with
averages $\mu_i = \frac12 + 4i\tau$. Let $S_{k/2+1},\ldots,S_k$ be sets of elements
with averages $\mu_{k/2+i} = \mu_i + \alpha_i \tau$ where the values 
$\alpha_1,\ldots,\alpha_{k/2} \in \{-1,1\}$ are chosen independently and uniformly at
random. Note that for every choice of $\alpha_i$'s, the minimal distances are all
$\eta_i = \tau$ for every $i=1,\ldots,k$.

Informally: the construction essentially ``gives away'' the values of $\mu_1,\ldots,\mu_{k/2}$
to the algorithm. But to satisfy the correct ordering property, the algorithm 
must distinguish between the cases where $\mu_{k/2+i} \in \{\mu_i \pm \tau\}$ for 
each of the values $i=1,\ldots,k/2$. We will argue that doing so with high probability
requires a large number of samples.

Let $A$ be any algorithm that satisfies the correct ordering property with probability 
at least $1-\delta$ on the class of inputs described above. For $i \in [k/2]$, let 
$q_i$ be the expected number of queries that $A$ makes to the values of elements in the 
set $S_{k/2+i}$.\footnote{Some clarification is needed: this expectation is over the 
internal randomness of $A$ and our choice of sets?} Let $\delta_i$ be the value that satisfies
$$
q_i = \frac{1}{8\tau^2} \ln( \frac1{16e\sqrt{\pi} \delta_i} ).
$$
By the (extension of the) Canetti--Even--Goldreich Theorem, the algorithm $A$ correctly
determines the order of $\mu_i$ and $\mu_{k/2+i}$ with probability at most $\delta_i$.

Since the choices of $\alpha_1,\ldots,\alpha_{k/2}$ are all made independently, the probability
that $A$ satisfies the correct ordering property is bounded above by
$(1-\delta_1)\cdots(1-\delta_{k/2})$. So, by the assumption on $A$ and the principle
of inclusion-exclusion, we have
$$
\delta \ge 1 - (1-\delta_1)\cdots(1-\delta_{k/2}) \ge \sum_{i=1}^{k/2} \delta_i - \sum_{i\neq j}
\delta_i \delta_j \ge \frac12 \sum_{i=1}^{k/2} \delta_i.
$$
(The last inequality holding when $\sum_i \delta_i \le 1/2$\ldots)
The total sample complexity of $A$ is
$$
\sum_{i=1}^{k/2} \frac{1}{8\tau^2} \ln( \frac1{16e\sqrt{\pi} \delta_i} )
= - \frac{k}{16\tau^2} \E_{i \in [k/2]} \ln( 16e\sqrt{\pi} \delta_i ).
$$
The function $-\ln(x)$ is convex, so by Jensen's inequality\cite{jensen}
\begin{align*}
- \frac{k}{16\tau^2} \E_{i \in [k/2]} \ln( 16e\sqrt{\pi} \delta_i )
&\ge - \frac{k}{16\tau^2} \ln( 16e\sqrt{\pi} \E_{i \in [k/2]} \delta_i ) \\
&\ge - \frac{k}{16\tau^2} \ln( 16e\sqrt{\pi} \cdot 2\delta/k ) \\
&= \Omega(\log(k/\delta) \frac{k}{2} \cdot \frac1{\tau^2}) \\
&= \Omega(\log(k/\delta) \sum_{i=1}^k 1/\eta_i^{\,2} ). 
\end{align*}
\end{proof}
}%techreport

\subsection{Discussion}\label{sec:discussion}

We now describe a few variations
of our algorithms. 

\techreport{
\stitle{Sampling with Replacement:}
Often, sampling with replacement is easier to implement
than sampling without replacement,
since we do not need to keep track of the
samples that have been taken.
On the other hand, 
sampling without replacement provides 
a smaller sample complexity,
since we only get ``fresh'' samples
every time.

If the algorithm does sampling with replacement
instead of without replacement, 
Serfling's inequality~\cite{serfling1974probability}
can be replaced with Hoeffding's inequality~\cite{hoeffding1963probability} simply by removing the
$(1-\frac{m/\kappa-1}N)$ term.

Thus, in the \ifocus\ algorithm,
we simply need to change Line 6 in the
algorithm to remove the  $(1-\frac{m/\kappa-1}{n_i})$
in the computation of the confidence interval.
As a result, the \ifocus\ algorithm for sampling
with replacement does not need to know
the values $n_1,$ $\ldots,$ $n_k$,
i.e., the number of elements in each group. 
\agpcomment{The sample complexity (Theorem~\ref{thm:analysis})
is modified by adding a ...}
}

\stitle{Visual Resolution Extension:}
Recall that in Section~\ref{sec:setup},
we discussed Problem~\ref{prob:relative-order-res},
wherein our goal is to only ensure that groups
whose true averages are sufficiently far enough to be correctly
ordered.
If the true averages of the groups are too close to each other,
then they cannot be distinguished on
a visual display, so 
expending resources resolving them is useless.

If we only require the correct ordering
condition to hold for groups whose 
true averages differ by 
more than some threshold $r$, we can
simply modify the algorithm to terminate 
once we reach a value of $m$ for which 
$\varepsilon_m < r/4$.
The sample complexity for this variant is 
essentially the same as in Theorem~\ref{thm:analysis} 
(apart from constant factors)
except that we replace each 
$\eta_i$ with $\eta_i^{(r)} = \max\{\eta_i, r\}$.

\stitle{Alternate Algorithm:}
The original algorithm we considered relies on
the standard and well-known
Chernoff-Hoeffding inequality~\cite{all-of-statistics}.
In essence, the algorithm---which we refer to as \iold, like \ifocus,
once again maintains confidence intervals
for groups, and stops sampling from inactive groups.
However, instead of taking one sample per iteration, \iold\ 
takes as many samples as necessary to divide the confidence interval
in two.
Thus, \iold\ is more aggressive than \ifocus.
% Note that we cannot avoid being this aggressive, since
% the Chernoff-Hoeffding bound (which is less specialized than
% the Hoeffding-Serfling bound) needs to be applied
% for every single round. 
\papertext{We provide the algorithm, the analysis,
and the pseudocode in our technical report~\cite{tr}.}
Needless to say, \iold, since it is so aggressive, 
ends up with a less desirable sample complexity than
\ifocus, and unlike \ifocus, \iold is not optimal.
We will consider \iold\ in our experiments.

\techreport{
\begin{algorithm}[h!]
\KwData{$i$, $\varepsilon$, $\delta$}
Draw $m = \frac{c^2}{2\varepsilon^2} \ln(2/\delta)$ samples $x_1,\ldots,x_m$ independently and 
uniformly at random from $S_i$\;
Return $\nu = \frac1t \sum_{j=1}^m x_j$\;
\caption{{\sc EstimateMean}}\label{alg:estimate-mean}
\end{algorithm}

\begin{algorithm}[h!]
\KwData{$S_1,\ldots,S_k,\delta$}
$\hat{\mu}_1, \ldots, \hat{\mu}_k \gets c/2$\; 
$\varepsilon_1, \ldots, \varepsilon_k \gets c/2$\;
$\delta_1, \ldots, \delta_k \gets 1/2k$\;
${\rm active}_1, \ldots, {\rm active}_k \gets {\rm True}$\;
\While{${\rm active}_1 \vee \cdots \vee {\rm active}_k$}{
  \For{$i=1,\ldots,k$}{
    \If {${\rm active}_i$}{
      Update $\varepsilon_i \gets \varepsilon_i / 2$ and $\delta_i \gets \delta_i / 2$\;
      Set $\hat{\mu}_i \gets$ {\sc EstimateMean}$(i,\varepsilon_i,\delta_i)$\;
      ${\rm active}_i \gets (\exists j \neq i : 
              [\hat{\mu}_i - \varepsilon_i, \hat{\mu}_i + \varepsilon_i] \cap 
              [\hat{\mu}_j - \varepsilon_j, \hat{\mu}_j + \varepsilon_j] \neq \emptyset)$\;
}
}
}
Return $\hat{\mu}_1,\ldots,\hat{\mu}_k$\;
\caption{\iold}\label{alg:old}
\end{algorithm}

The algorithm is listed in Algorithm~\ref{alg:old}, and uses
the subroutine in Algorithm~\ref{alg:estimate-mean}.

\begin{theorem}
\label{thm:basic}
With probability at least $1-\delta$, the values $\overline{\mu}_1,\ldots,\overline{\mu}_k$ returned
by the \iold\ algorithm satisfy $\overline{\mu}_i < \overline{\mu}_j$ iff 
$\mu_i < \mu_j$ for every $1 \le i < j \le k$ and this result is obtained after making at most
$O( \log(k/\delta) \sum_{i=1}^k \frac{\log(1/\eta_i)}{\eta_i^{\,2}} )$ queries.
\end{theorem}

The proof of the theorem relies on the lemma about the estimated means established via the
Chernoff--Hoeffding bound.

\begin{lemma}
\label{lem:estimate}
For any $0 < \varepsilon, \delta < 1$, $O(\frac{1}{\varepsilon^2} \log(1/\delta))$ samples drawn 
uniformly at random (with replacement) from $S_i$ suffice to obtain an estimate $\nu_i$ of
$\mu_i$ that satisfies $\mu_i - \varepsilon \le \nu_i \le \mu_i + \varepsilon$ with probability at
least $1 - \delta$.
\end{lemma}

We're now ready to complete the analysis of the algorithm.

\begin{proof}[Theorem~\ref{thm:basic}]
By the union bound, with probability at least $1-\delta$
at every execution of the inner loop of the
{\sf IterativeRefinement} algorithm, we have $\overline{\mu_i} \in [\mu_i - \varepsilon_i,
\mu_i + \varepsilon_i]$. For the rest of the analysis, assume that this condition holds.

The correctness of the algorithm follows directly from the fact that it stops refining
the estimate $\overline{\mu}_i$ only when the confidence interval around it is disjoint
from the intervals around its other estimates $\overline{\mu}_j$ for every $j \neq i$.

To establish the query complexity of the algorithm, we first note that the algorithm
stops refining the estimate $\overline{\mu}_i$ whenever $\varepsilon_i < \eta_i/2$. This is
because at this point, all the confidence intervals for the estimates of $\mu_j$, $j \neq i$,
that are still active have length less than $\eta_i / 2$: since $\eta_i$ measures the minimal
distance between $\mu_i$ and any other $\mu_j$, these intervals cannot intersect. 
Therefore, by Lemma~\ref{lem:estimate}, at most 
$O( \log(k/\delta) \frac{\log(1/\eta_i)}{\eta_i^{\,2}})$ samples are required to estimate
$\overline{\mu}_i$ before active$_i$ is set to false.
\end{proof}
}%techreport

\techreport{
\stitle{Theory Remarks:}
We now state some additional remarks regarding
the theorems that we have used to derive 
correctness and sample complexity of \ifocus.
\begin{remark}
The original statement of Serfling's inequality (1974) is for the value
$S_n/n$ instead of $\max_{1 \le k \le n} S_k/k$. 
See McDiarmid (\S2 of {\it Concentrations}, 1998) for a discussion on how
this and other bounds obtained via Bernstein's elementary inequality
($\Pr[ Z \ge t] \le e^{-ht}\E[e^{hZ}]$) can all be extended to 
maxima.
See also Bardenet and Maillard (2013) for a discussion of the maximal
version of the Hoeffding--Serfling inequality and the following slight sharpening
of the inequality for the case where $n \ge N/2$.
\end{remark}

\begin{remark}
Actually, Serfling's inequality (1974) is also stated as a one-sided inequality
bounding the probability that $S_n/n - \mu$ is greater than $\varepsilon$. 
The two-sided inequality is obtained by applying the same inequality to the
sum of the random variables $Y_i = 1 - X_i$.
\end{remark}

\begin{remark}
The argument of Theorem~\ref{thm:hs} is essentially an adaptation of the 
upper bound argument in the proof of the 
Law of the Iterated Logarithm. See, e.g., Ledoux and Talagrand (1991) for details.
\end{remark}
}%techreport

%!TEX root=main.tex

\section{System Description}\label{sec:system}

We evaluated our algorithms on top of a new database system
we are building, called \ntail, that is designed to produce a random sample of
records matching a set of ad-hoc conditions.
%. \ntail is a system that is tailored
%towards a new database interaction paradigm, {\em browsing}. Given a
%query, \ntail returns a small (in comparison with the query result)
%randomly chosen number (i.e., a ``screenful'') of records that satisfy
%the query conditions as quickly as possible, independent of the total
%number of records in the query result. While \ntail can be used in a
%browsing scenario, it can also be used in a {\em sampling} scenario,
%for example, as the back-end for our visualization system.
%To demonstrate the actual (wall clock) performance
%of our algorithms, we implemented them in a new database system we are building for operating on samples, called
%\ntail.
%\ntail\ is a system that is tailored towards 
%a new database interaction paradigm, {\em browsing}.
%Given a query, \ntail\ returns a small (in comparison with
%the query result) randomly chosen number (i.e., a ``screenful'') of records that satisfy the query conditions 
%as quickly as possible, independent of the total number
%of records in the query result.
%While \ntail\ can be used in a browsing scenario,
%it can also be used in a {\em sampling} scenario,
%for example, as the back-end for our visualization system.
To quickly retrieve satisfying tuples, \ntail\
uses in-memory bitmap-based indexes. 
We refer the reader to the demonstration paper for
the full description of \ntail's bitmap index
optimizations~\cite{ntail}.
% \alkim{Guessing we need to change this to
% tech report now?})
Traditional in-memory bitmap indexes allow rapid retrieval 
of records matching ad-hoc user-specified predicates. 
In short, for every value of every attribute in the relation
that is indexed, the bitmap index
records a 1 at location $i$ when the $i$th tuple matches the value for that attribute,
or a 0 when the tuple does not match that value for that attribute.
While recording this much information for every value of 
every attribute could be quite costly, in practice,
bitmap indexes can be compressed significantly, enabling
us to store them very compactly in memory~\cite{DBLP:conf/cikm/Koudas00,
DBLP:journals/tods/WuOS06, DBLP:journals/tods/WuSS10}.
\ntail employs several other optimizations to 
store and operate on these bitmap indexes very efficiently.
%\alkim{Possibly get rid of this line:
%In our example scenario in the introduction,
%\ntail's indexes can be used to retrieve, at any point during execution,
%an additional random sample specifically from group {\tt AA} or {\tt JB}.
%}
{\em Overall, \ntail's in-memory bitmap indexes allow it to retrieve and 
return a tuple from disk matching certain conditions in constant time.}
\agpneutral{Note that even if the bitmap is dense or sparse, the guarantee
of constant time continues to hold because the bitmaps are organized in
a hierarchical manner (hence the time taken is logarithmic in the total
number of records or equivalently the depth of the tree).
}
\agpneutral{\ntail can be used in two modes: either a column-store
or a row-store mode. For the purpose of this paper, 
we use the row-store configuration, enabling us to eliminate
any gains originating from the column-store.}
\ntail\ is written in C++ and uses the Boost library
for its bitmap and hash map implementations.

% \squishlist
% \item 
% Goal: Support Browsing and Sampling queries 
% (give me any $k$ that satisfies a predicate), 
% without knowing the form of the queries in advance
% \item 
% Core Technology: Fusion IO + Bitmap-based
% Index search. 
% \item 
% Architecture: Refer to a diagram and 
% talk about the architecture
% \item 
% Supports: Any selection query on a star-schema.
% \item 
% Challenges: Designing a collection of Bitmap 
% Indices that support retrieval
% with as few false positives as possible. 
% (Not the focus of this paper.)
% Here, we assume that there is a bitmap index
% for each attribute value of each attribute.
% \agp{most general description.}
% \squishend

%!TEX root=main.tex

\begin{figure*}[!t]
\vspace{-15pt}
\centering
\subfigure{
\includegraphics[height=1.8in]{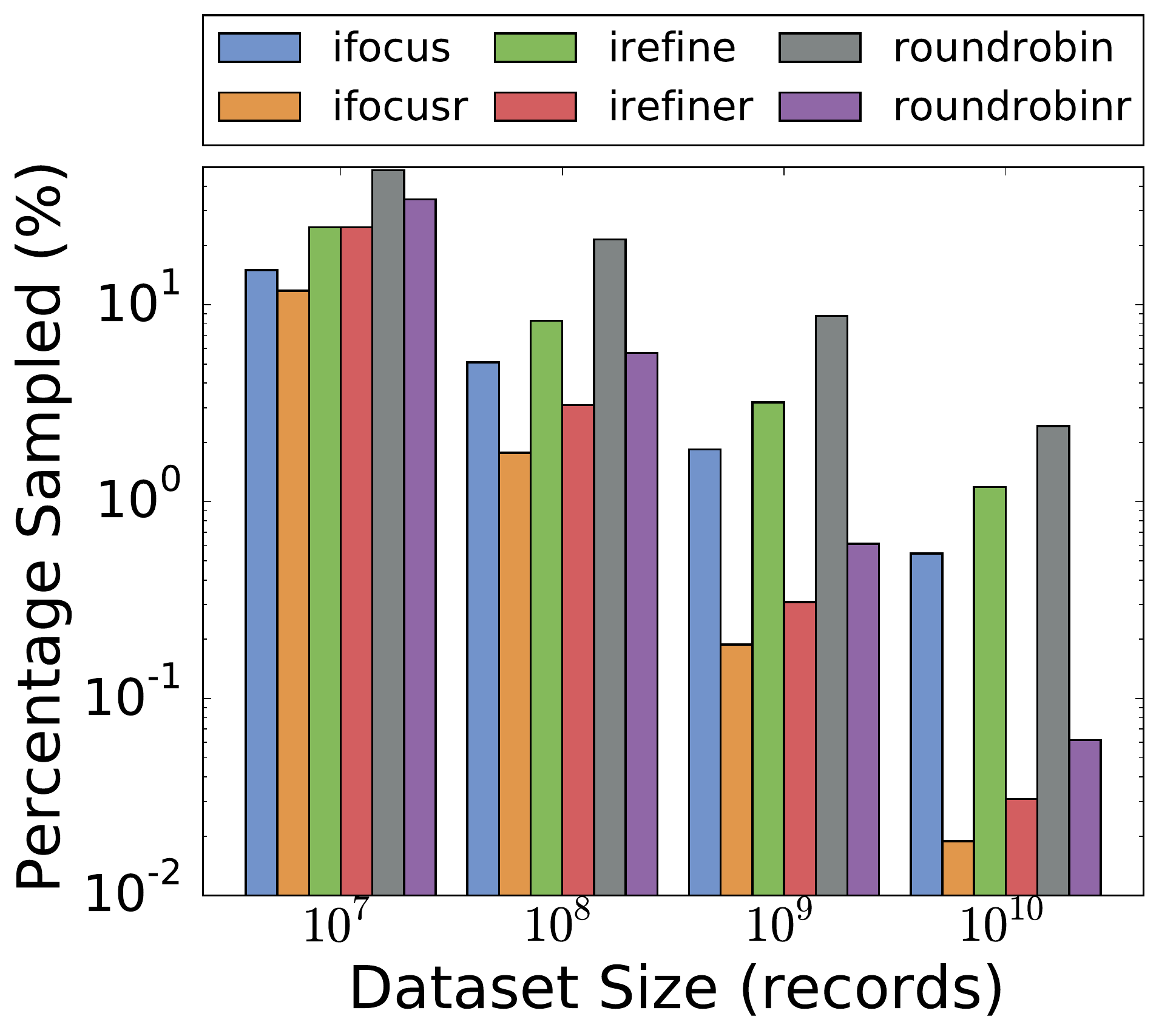}
\label{fig:data_size.sampled}
}
%\hspace{5pt}
\subfigure{
\includegraphics[height=1.8in]{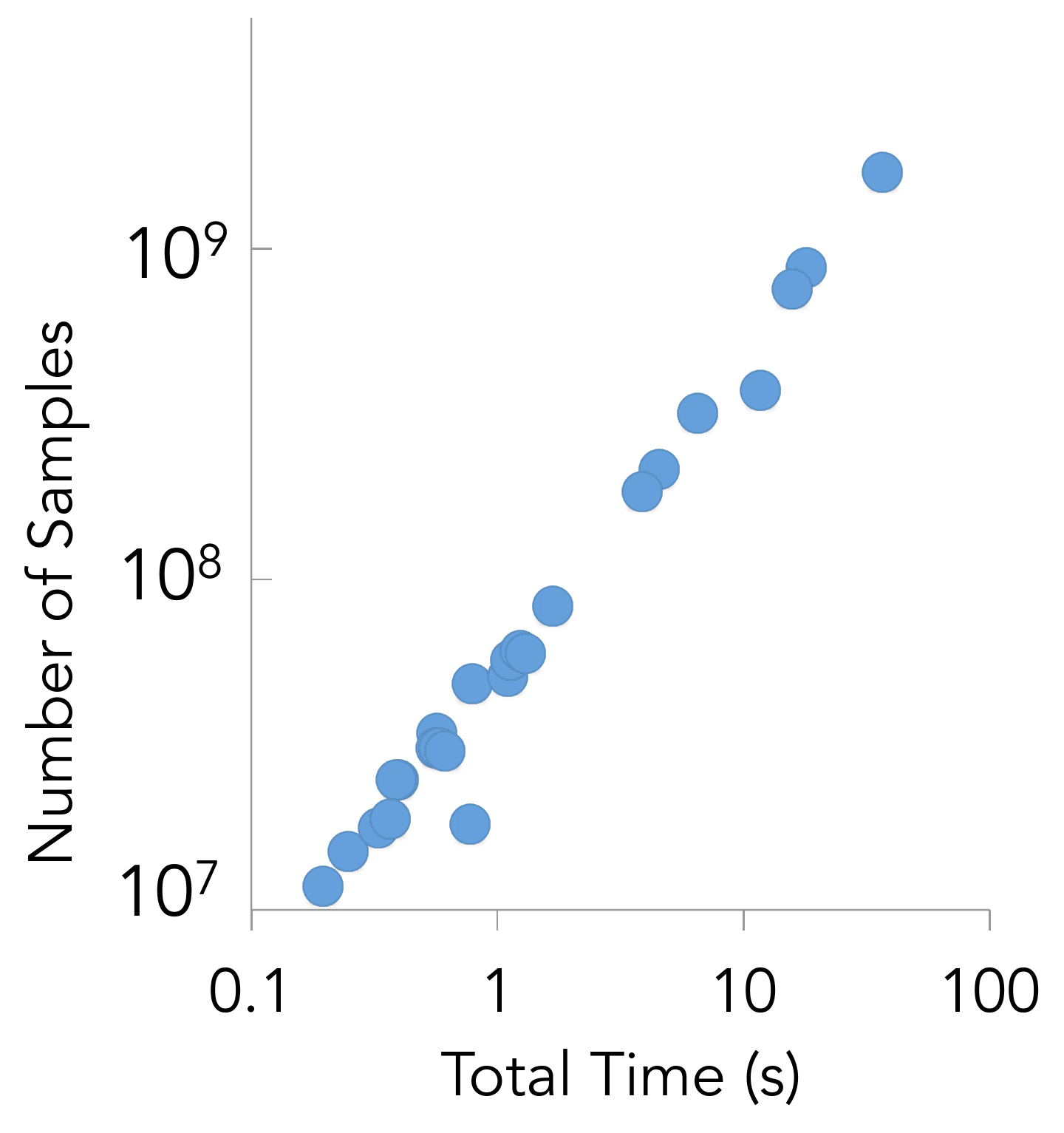}
\label{fig:sample-to-runtime}
}
%\hspace{5pt}
\subfigure{
\includegraphics[height=1.8in]{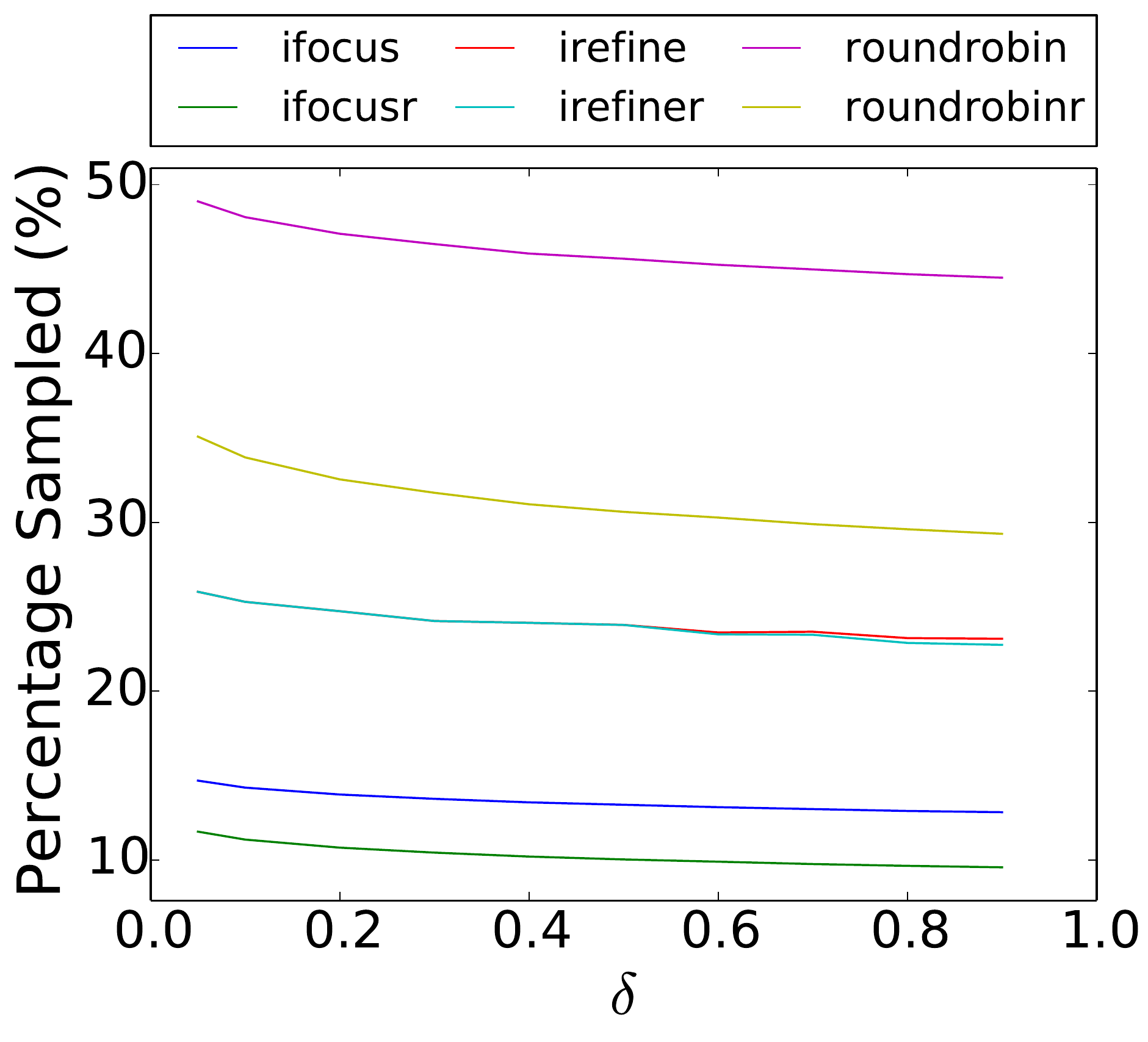}
\label{fig:mixture.mean.s}
}
\vspace{-13pt}
\caption{(a)Impact of data size (b) Scatter plot of samples vs runtime (c) Impact of $\delta$}
\vspace{-15pt}
\end{figure*}

\begin{figure*}[!t]
%\vspace{-15pt}
\centering
\subfigure{
\includegraphics[height=2in]{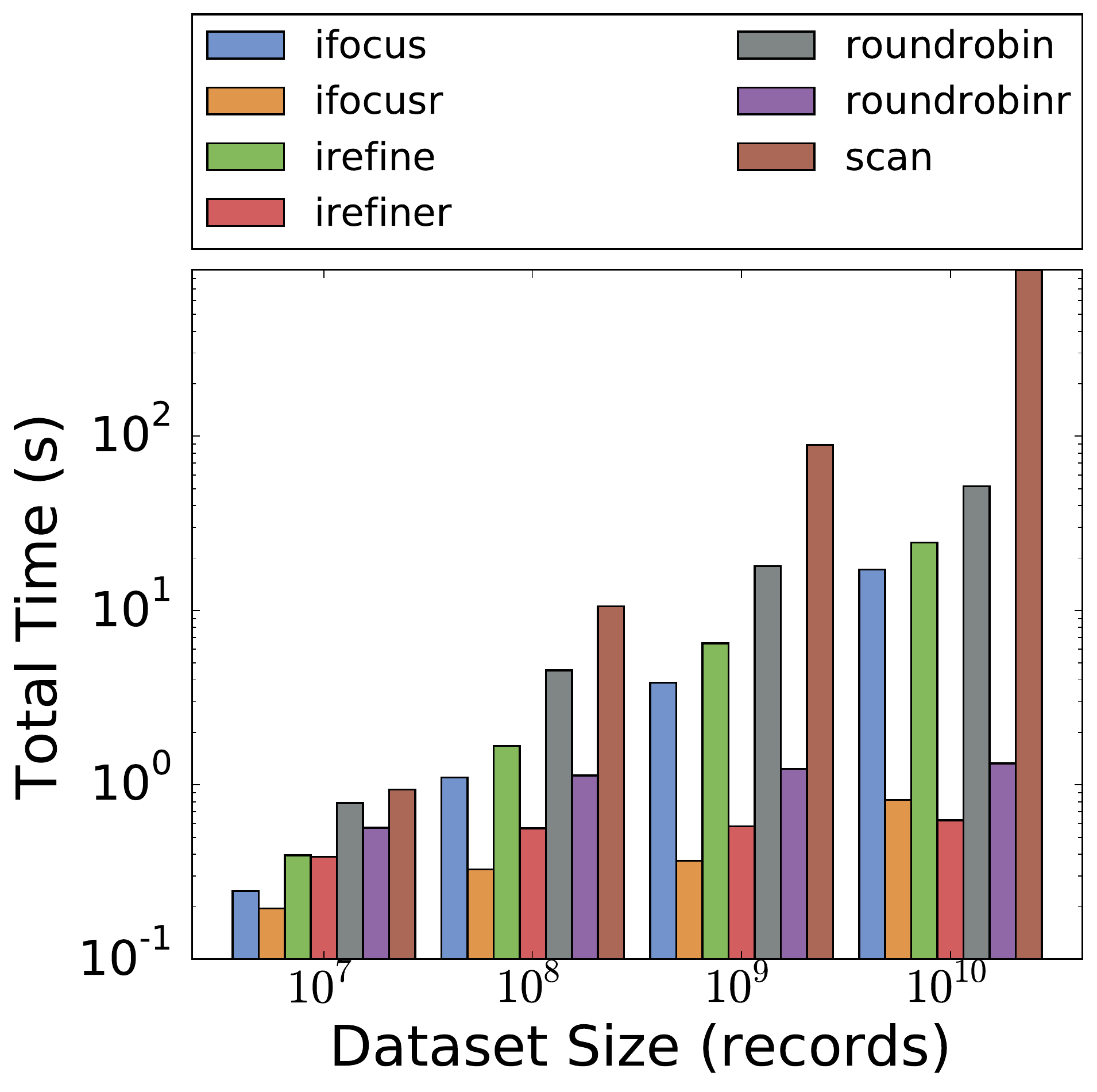}
\label{fig:data_size.total_time}
}
%\hspace{5pt}
\subfigure{
\includegraphics[height=2in]{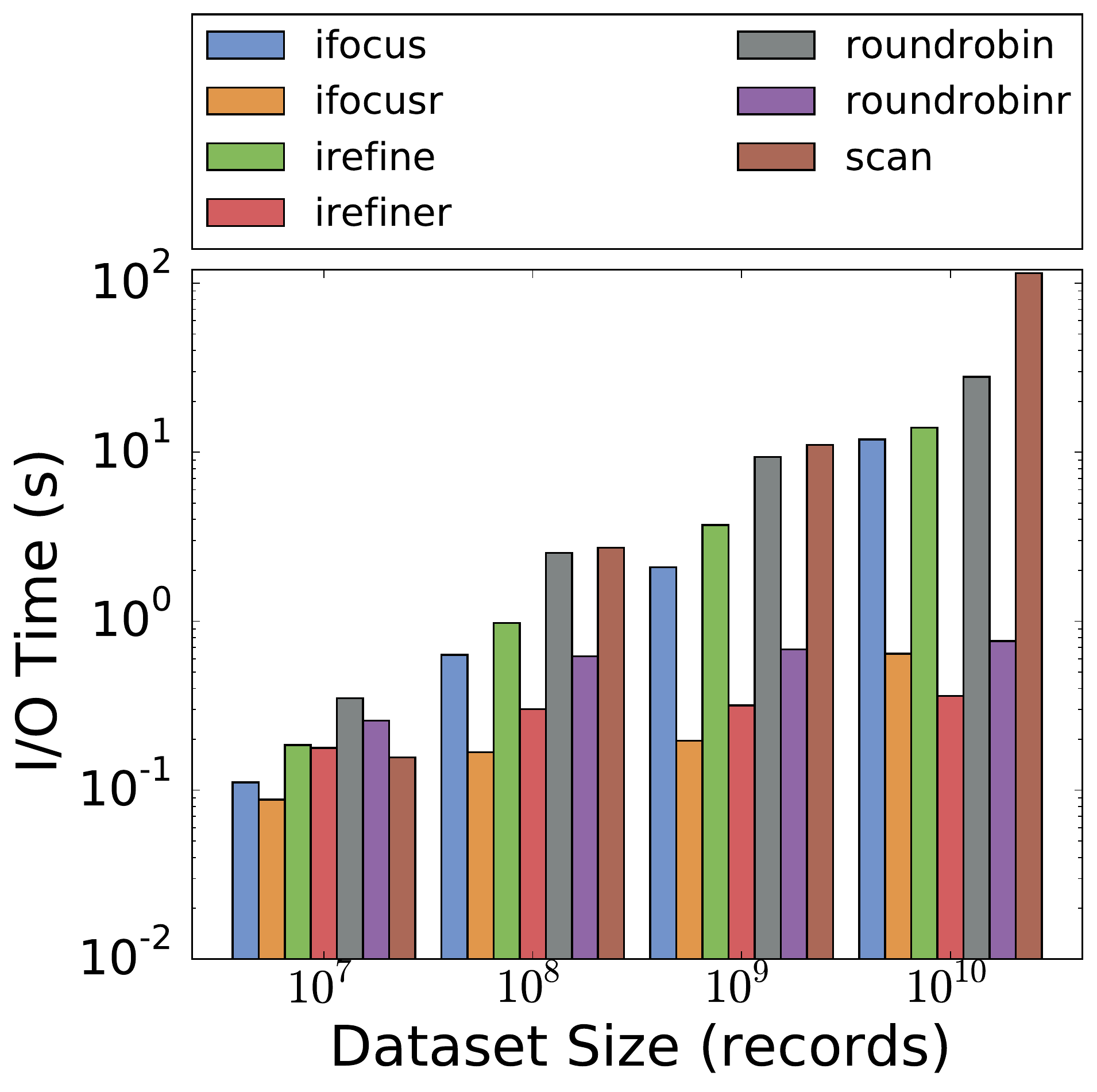}
\label{fig:data_size.io_time}
}
%\hspace{5pt}
\subfigure{
\includegraphics[height=2in]{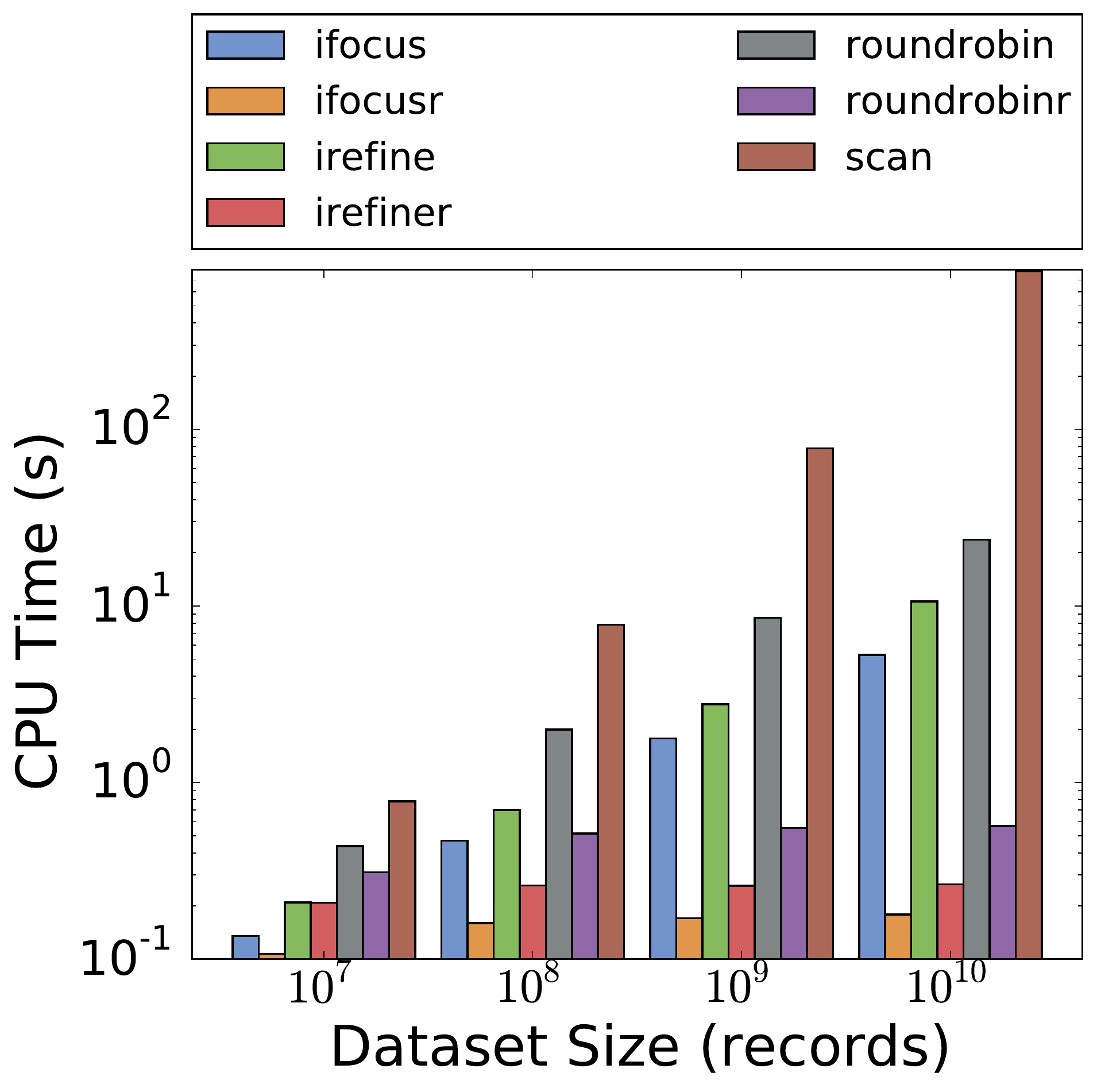}
\label{fig:data_size.cpu_time}
}
\vspace{-10pt}
\caption{(a)Total time vs dataset size (b) I/O time vs dataset size (c) CPU time vs dataset size}
\vspace{-15pt}
\end{figure*}

\section{Experiments}\label{sec:exp}

%In this section, we experimentally evaluate our algorithms
%versus other algorithms on a variety of 
%synthetic and real-world datasets.

% \alkim{I collapsed the goals from the next section into the introduction
%   here to save space, and because I felt like that it served better as
%   an overview paragraph rather than a goals paragraph.
% }
In this section, we experimentally evaluate our algorithms
versus traditional sampling techniques on a variety of 
synthetic and real-world datasets.
We evaluate the algorithms on three different metrics: the number of
samples required (sample complexity), the accuracy of the produced
results, and the wall-clock runtime performance on our prototype sampling
system, \ntail.

\subsection{Experimental Setup}

%\stitle{Goals:}
%The primary goal of our experimental study is to evaluate the number 
%of samples required by our algorithm versus traditional sampling techniques (sample complexity).
%We also measure the accuracy of the produced results,
%to test whether the algorithms respect the correct ordering property.
%Finally, we study the runtime performance of our algorithms by measuring
%the wall time of their execution on our prototype sampling system,
%\ntail.

\stitle{Algorithms:}
Each of the algorithms we evaluate takes as a 
parameter $\delta$, a bound on the probability that the algorithm
returns results that do not obey the ordering property.
That is, all the algorithms are guaranteed to return
results ordered correctly with probability $1-\delta$, no
matter what the data distribution is.

The algorithms are as follows:

\begin{asparaitem}
\item{\bf \ifocus($\delta$)}: In each round, this algorithm takes
an additional sample from all active groups, ensuring
that the eventual output has accuracy greater than $1- \delta$,
as described in Section~\ref{sec:alg-description}.
This algorithm is our solution for Problem~\ref{prob:relative-order}.

\item{\bf \ifocusr($\delta, r$)}: In each round, this algorithm takes
an additional sample from all active groups, ensuring
that the eventual output has accuracy greater than $1- \delta$,
for a relaxed condition of accuracy based on resolution.
Thus, this algorithm is the same as the previous, except 
that we stop at the granularity of the resolution value.
This algorithm is our solution for Problem~\ref{prob:relative-order-res}.

\item{\bf \iold($\delta$)}: In each round, this algorithm divides
all confidence intervals by half for
all active groups, ensuring that the eventual output has
accuracy greater than $1- \delta$,
as described in Section~\ref{sec:discussion}.
Since the algorithm is aggressive in taking samples to divide
the confidence interval by half each time, we expect it
to do worse than \ifocus.

\item{\bf \ioldr($\delta, r$)}: This is the \iold algorithm except we
  relax accuracy based on resolution as we did in \ifocusr.
%  In each round, this algorithm divides
%all confidence intervals by half for
%all active groups, ensuring that the eventual output has
%accuracy greater than $1- \delta$, for a relaxed resolution-based accuracy condition.
%This algorithm is the same as the previous, except that we
%stop at the granularity of the resolution.
\end{asparaitem}

\noindent We compare our algorithms against the following baseline:

\begin{asparaitem}
\item{\bf \ir($\delta$)}: In each round, this algorithm takes
an additional sample from all groups, ensuring
that the eventual output respects the order with 
probability than $1- \delta$.
This algorithm is similar to conventional stratified sampling
schemes~\cite{casella}, 
except that the algorithm 
has the guarantee that the ordering property
is met with probability greater than $1 - \delta$.
We adapted this from existing techniques to
ensure that the ordering property
is met with probability greater than $1-\delta$.
We cannot leverage any pre-existing techniques
since they do not provide the desired guarantee.

\end{asparaitem}

%\noindent To study the impact of the relaxed output condition 
%based on resolution, we also consider the following algorithm:

\begin{asparaitem}
\item{\bf \irr($\delta, r$)}: This is the \ir algorithm except we relax
  accuracy based on resolution as we did in \ifocusr.
%  In each round, this algorithm takes
%an additional sample from all groups, ensuring
%that the eventual output has accuracy greater than $1- \delta$,
%for a relaxed condition of accuracy based on resolution.
%This algorithm is the same as \ir, except that we stop earlier
%when the estimates for two groups are close enough to each other.
%We include this algorithm to study the impact of the relaxed output
%condition based on resolution.

\end{asparaitem}

\stitle{System:}
We evaluate the runtime performance of all our algorithms on our
early-stage \ntail prototype. We measure both the CPU times and the I/O
times in our experiments to definitively show that our improvements are
fundamentally due to the algorithms rather than skilled engineering. In
addition to our algorithms, we implement a \scan operation in \ntail,
which performs a sequential scan of the dataset to find the true means
for the groups in the visualization. The \scan operation represents an
approach that a more traditional system, such as PostgreSQL, would take
to solve the visualization problem. Since we have both our sampling
algorithms and \scan implemented in \ntail, we may directly compare these two approaches. 
We ran all experiments on a
64-core Intel Xeon E7-4830 server running Ubuntu 12.04 LTS; however, all
our experiments were single-threaded. We use 1MB blocks to read from
disk, and all I/O is done using Direct I/O to avoid any speedups we
would get from the file buffer cache.
\papertext{Note that our \ntail system is still in its early stages and under-optimized --- 
we expect our performance numbers to only get better as the system improves.}
\techreport{Finally, we would like to state that our \ntail prototype is still in
its early stages. The current implementation runs on top of a basic
bitmap/hash map implementation with little-to-no optimization code. We
believe custom writing our own bitmap/hash map implementations could
lead to substantial improvements in the CPU overhead. Parallelization is
another way to get significant gains in performance. We can easily
parallelize our sampling workload to fully utilize the I/O bandwidth
since random sampling tends to be an independent operation. In short,
with a few optimizations, we fully expect runtime performance of our
algorithms to be even better than the results we present in the
following sections.}

\stitle{Key Takeaways:}
Here are our key results from experiments in 
Sections~\ref{sec:synth-exp} and \ref{sec:real-exp}:
\vspace{-7pt}
\begin{framed}
\vspace{-5pt}
\noindent (1) Our \ifocus\ and \ifocusr\ ($r$=1\%) 
algorithms yield 
\begin{denselist} 
\item up to 80\% and 98\%  reduction in sampling and
  79\% and 92\% in runtime (respectively)
as compared to \ir,
on average, across a range of very large synthetic datasets, and

\item up to 70\% and 85\% reduction in runtime (respectively)
as compared to \ir, for multiple attributes in a realistic, large flight 
records dataset~\cite{flight-data}.

\end{denselist}
\vspace{0.5em}
\noindent (2) The results of our algorithms (in all of the experiments we have conducted)
always respect the correct ordering property.

\vspace{-5pt}
\end{framed}
\vspace{-7pt}

\subsection{Synthetic Experiments}\label{sec:synth-exp}
We begin by considering experiments on synthetic data. 
The datasets we ran our experiments on are as follows:
\begin{asparaitem}
\techreport{
\item{\bf Truncated Normals (truncnorm):} For each group, we select
a mean $\sigma$ sampled uniformly at random from $[0,100]$, and select
a variance $\Delta$ from $\{4, 25, 64, 100\}$.  We
generate values from a normal distribution
for each group using the selected values. 
These normals are truncated at 0 and 100 
to ensure that the values are bounded.
}

\item{\bf Mixture of Truncated Normals (mixture):} For each group, we select
a collection of normal distributions, in the following way:
we select a number sampled at random from $\{1, 2, 3, 4, 5\}$, indicating
the number of truncated normal distributions that comprise each group.
For each of these truncated normal distributions, we select
a mean $\sigma$ sampled at random from $[0, 100]$, and a
variance $\Delta$ sampled at random from $[1, 10]$. 
We repeat this for each group.

\techreport{
\item{\bf Bernoulli (bernoulli):} For each group, we select
a mean sampled at random from $[0, 100]$. 
Then, we construct our group by
sampling between two values $\{0, 100\}$
with the bias equal to the chosen mean.
Thus, this is a Bernoulli distribution~\cite{all-of-statistics} with a mean
selected up front.
Note that this distribution has higher variance than
those that are normal.
% We chose this distribution because it is possibly the most
% ``extreme'' distribution we could study. \srm{I don't know what you mean by extreme!}
}
\item{\bf Hard Bernoulli (hard):} Given a parameter $\gamma<2$, we fix the mean
for group $i$ to be $40 + \gamma \times i$, and then
construct each group by sampling between two values
$\{0, 100\}$ with bias equal to the mean.
Note that in this case,
$\eta$, the smallest distance between two means, is equal to $\gamma$.
Recall that $c^2/\eta^2$ is a proxy for how difficult the input instance is
(and therefore, how many samples need to be taken).
We study this scenario so that we can control the difficulty of 
the input instance.
\end{asparaitem}

Our default setup consists
of $k = 10$ groups, with $10M$ records in total,
equally distributed across all the groups, with $\delta = 0.05$
(the failure probability) and $r = 1$.
Each data-point is generated by repeating the experiment
100 times.
That is, we construct 100 different datasets
with each parameter value, and measure
the number of samples taken when the 
algorithms terminate, whether
the output respects the correct ordering property, and the CPU and I/O
times taken by the algorithms.
For the algorithms ending in {\sc R}, i.e., 
those designed for a more relaxed property 
leveraging resolution, we check if the output
respects the relaxed property rather than the more
stringent property.
We focus on the mixture
distribution for most of the experimental
results, since we expect it to be the most representative
of real world situations, using the hard Bernoulli in a few cases.
We have conducted extensive experiments with other distributions as well,
and the results are similar.
\techreport{We will describe these experiments whenever the behavior for those 
distributions is significantly different.}
\papertext{The complete
experimental results can be found in 
our technical report~\cite{tr}.}

\stitle{Variation of Sampling and Runtime with Data Size:} 
We begin by measuring the sample complexity and wall-clock times of our
algorithms as the data set size varies.
\vspace{-7pt}
\begin{framed}
\vspace{-5pt}
\noindent {\em Summary:} Across a variety of dataset sizes, our algorithm
\ifocusr\ (respectively \ifocus) 
performs better 
on sample complexity and runtime than \ioldr\ (resp.~\iold) 
which performs significantly better than \irr\ (resp.~\ir).
Further, the resolution improvement versions take many fewer samples than 
the ones without the improvement. In fact, for any dataset size greater
than $10^8$, the resolution improvement versions take a constant number
of samples and still produce correct visualizations.
\vspace{-5pt}
\end{framed}
\vspace{-7pt}

\begin{figure*}[!t]
\centering
\subfigure{
\includegraphics[height=1.5in]{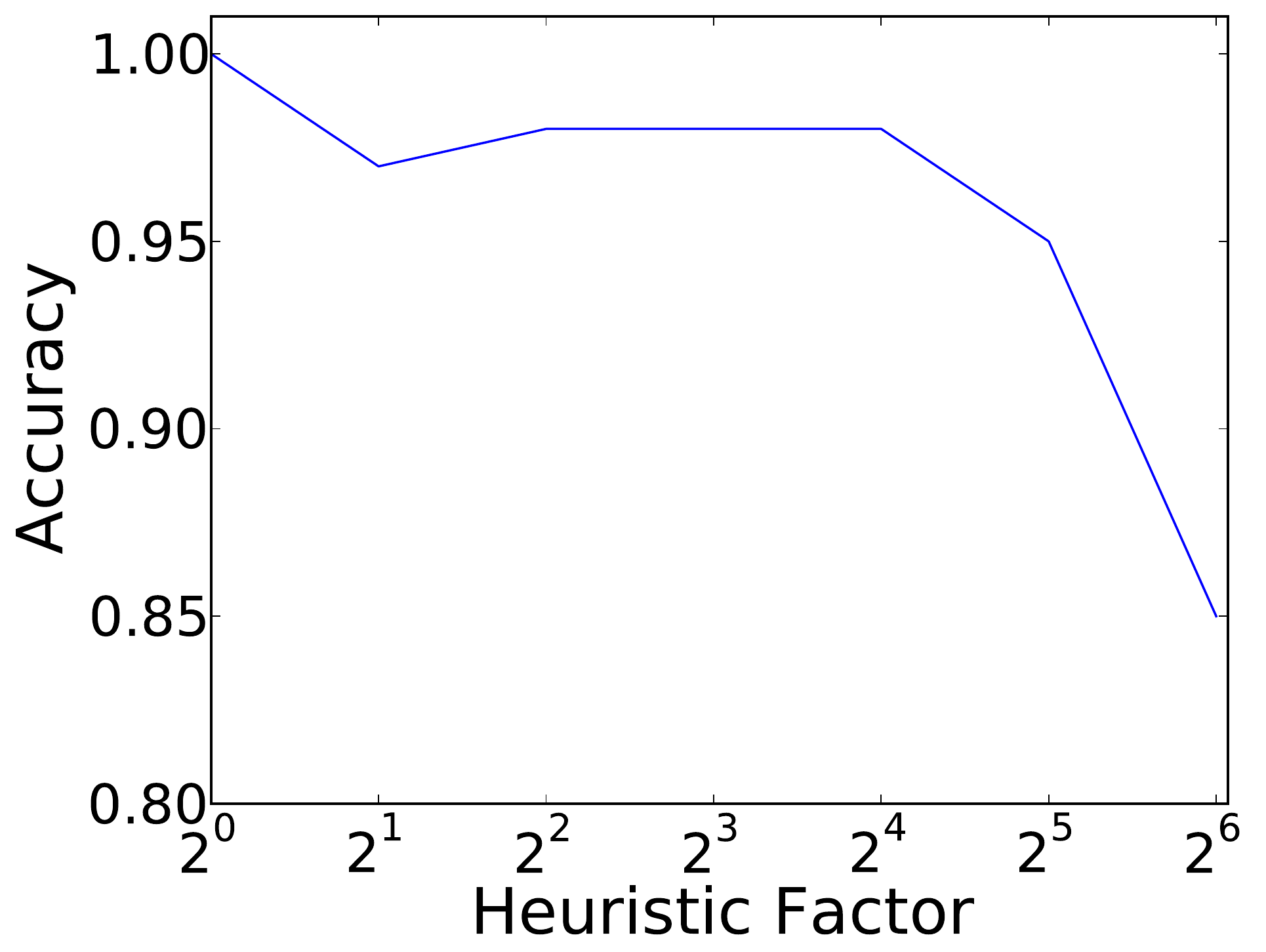}
\label{fig:mixture.factor.a}
}
%\hspace{-5pt}
\subfigure{
\includegraphics[height=1.5in]{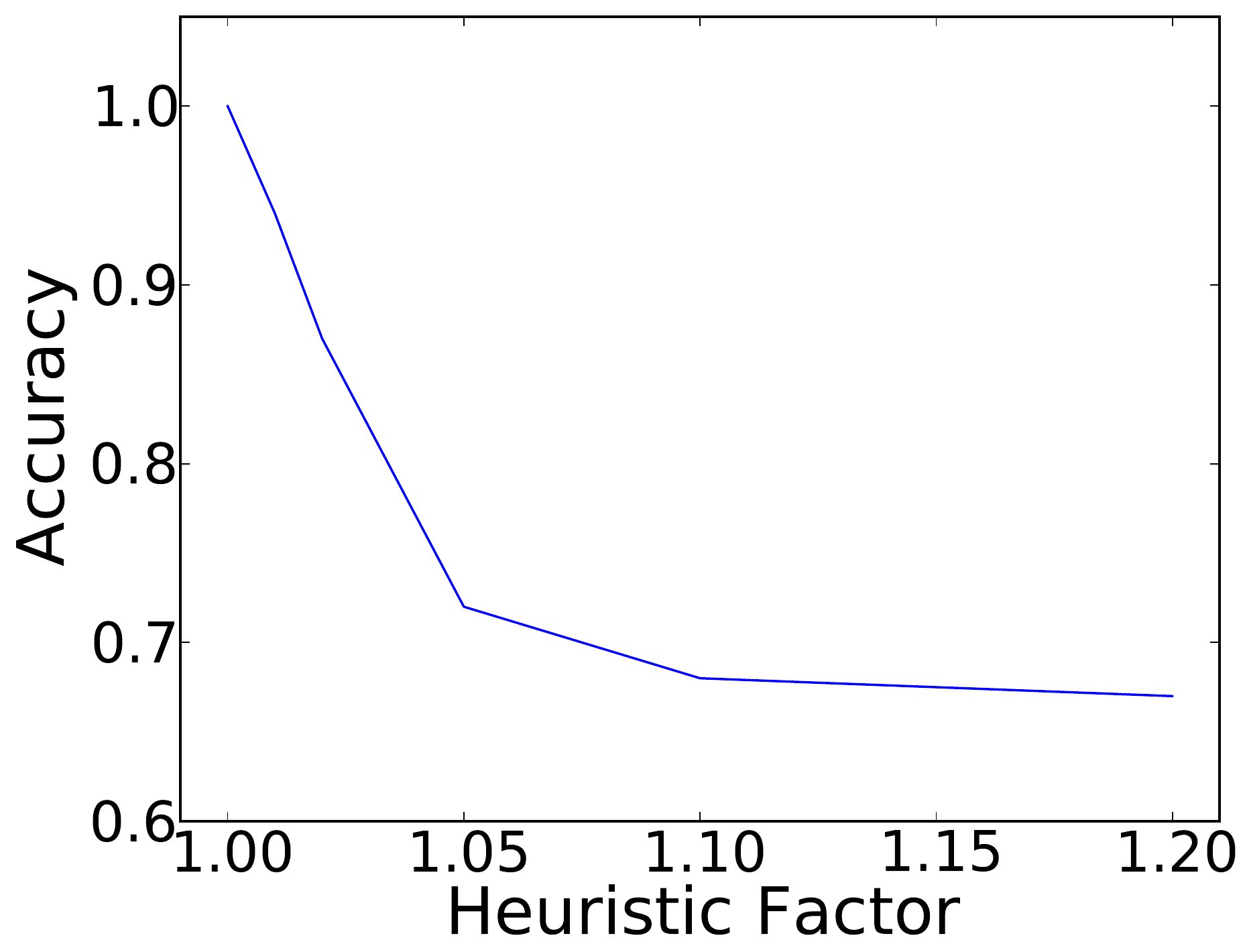}
\label{fig:hard.0.1.a}
}
%\hspace{-5pt}
\subfigure{
\includegraphics[height=1.5in]{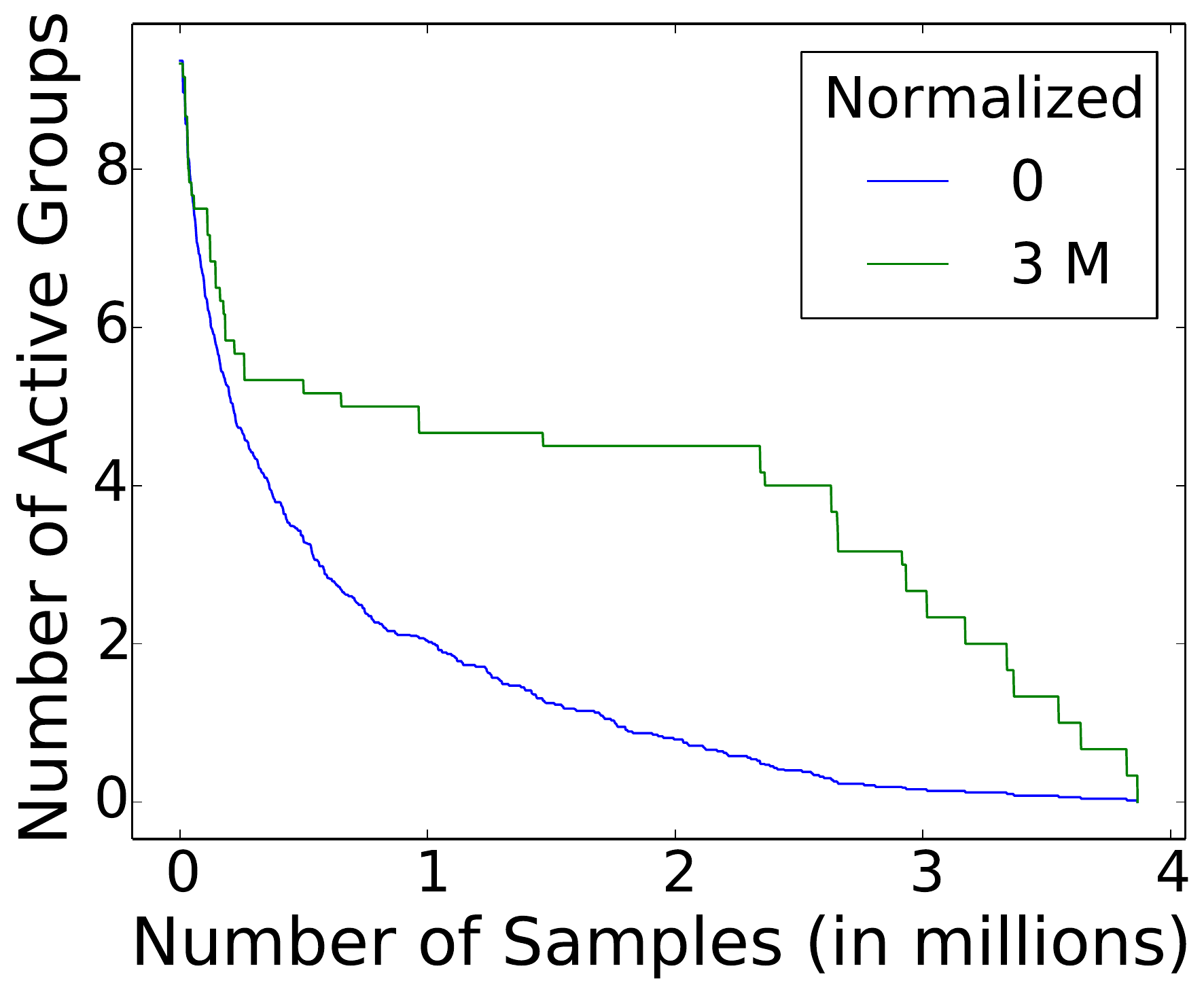}
\label{fig:mixture.active_intervals.c}
}
\vspace{-10pt}
\caption{(a) Impact of heuristic shrinking factor on accuracy 
(b) Impact of heuristic shrinking factor for a harder case
(c) Studying the number of active intervals as computation proceeds
}
\vspace{-15pt}
\end{figure*}

\agpneutral{
Figure~\ref{fig:data_size.sampled} shows the percentage of the dataset
sampled on average as a function of dataset size (i.e., total number of
tuples in the dataset across all groups) for the six algorithms above.
The data size ranges from $10^7$ records to $10^{10}$ records (hundreds of GB). 
%We postulate that
%10B records (which is on the order of hundreds of GBs) is the maximum
%number of records a person would wish to store on a single node.
%Figure~\ref{fig:data_size.raw_sampled} depicts the same data as the absolute number of
%records sampled instead of sample percentages.
Note that the figure
is in log scale.
}
%Figure~\ref{fig:mixture.data_size.mean} shows the absolute percentage of the dataset sampled
%on average
%as a function of dataset size (i.e., total number of tuples in the
%dataset across all groups) for the six
%algorithms described above.
%Figure~\ref{fig:mixture.data_size.ratio.mean} shows the the ratio between the sampling percentage of select pairs
%of algorithms as a function of dataset size.

\agpneutral{
%Consider Figure~\ref{fig:data_size.sampled}, for dataset size = $10^7$.
Consider the case when dataset size = $10^7$ in
Figure~\ref{fig:data_size.sampled}.
Here 
\ir\ samples $\approx$50\% of the data, while \irr\ samples around 35\% of the 
dataset. On the other hand, our \iold\ and \ioldr\ algorithms
both sample around 25\% of the dataset,
while \ifocus samples around 15\% and \ifocusr around 10\% of the dataset.
Thus, compared to the vanilla \ir\ scheme, all our algorithms reduce the number of samples 
required to reach the order guarantee, by up to $3\times$.
This is because our algorithms focus on the groups that are actually contentious,
rather than sampling from all groups uniformly.
}

\agpneutral{
As we increase the dataset size, we see that the sample percentage
decreases almost linearly for our algorithms, suggesting that there
is some fundamental upper bound to the number of samples required, 
confirming Theorem~\ref{thm:analysis}. With resolution improvement, this
upper bound becomes even more apparent. In fact, we find that the raw
number of records sampled for \ifocusr, \ioldr, and \irr all remained
constant for dataset sizes greater or equal to $10^8$. In addition, as
expected,
\ifocusr (and \ifocus) continue to outperform all other algorithms at
all dataset sizes.
}

\agpneutral{
The wall-clock total, I/O, and CPU times for our algorithms running on
\ntail can be found in Figures \ref{fig:data_size.total_time},
\ref{fig:data_size.io_time}, and \ref{fig:data_size.cpu_time},
respectively, also in log scale.
Figure~\ref{fig:data_size.total_time} shows that for a dataset of size
of $10^{9}$ records (8GB), \ifocus/\ifocusr take 3.9/0.37 seconds to
complete, \iold/\ioldr take 6.5/0.58 seconds to complete, \ir/\irr take
18/1.2 seconds to complete, and \scan takes 89 seconds to
complete. This means that \ifocus/\ifocusr has a 23$\times$ speedup and
241$\times$ speedup relative to \scan in producing accurate visualizations.
}

\agpneutral{
As the dataset size grows, the runtimes for the sampling
algorithms also grow, but sublinearly, in accordance
to the sample complexities. In fact, as alluded earlier, we see that the
run times for \ifocusr, \ioldr, and \irr are nearly constant for all
dataset sizes greater than $10^8$ records. There is some variation, e.g.,
in I/O times for \ifocusr at $10^{10}$ records, which we believe is due to random noise.
 In contrast,
 \scan yields linear scaling,
leading to unusably long wall-clock (i.e., 898 seconds at $10^{10}$
records.)
}

\agpneutral{
We note that not only does 
\ifocus beat out \ir, and \ir beat out \scan
for every dataset size in total time, but this remains true for both
I/O and CPU time as well. Sample complexities explain why \ifocus
should beat \ir.  It is more surprising that
\ifocus, which uses random I/O, outperforms \scan, which only uses sequential
I/O.  {\em The answer is that so few samples are required
the cost of additional random I/O is exceeded by the additional scan time;
 this becomes more true as the dataset
size increases.} As for CPU time, it
highly correlated with the number of samples, so algorithms that
operate on a smaller number of records outperform algorithms
that need more samples.
}

\agpneutral{
The reason that CPU time for \scan is actually greater than the I/O time is that for every record read, it must
update the mean and the count in a hash map keyed on the group.  While
Boost's ${\tt unordered\_map}$ implementation is very efficient, our disk subsystem is able to read about 800 MB/sec, and
a single thread on our machine can only perform about 10 M hash probes and updates / sec.
However, even if we discount the CPU overhead of \scan, we
find that total wall-clock time for \ifocus and \ifocusr is at least an
{\em order of magnitude better than just the I/O time} for \scan. For $10^{10}$ records,
compared to \scan's 114 seconds of sequential I/O time, \ifocus has a
total runtime of 13 seconds, and \ifocusr has a total runtime in 0.78
seconds, giving a speedup of at least 146$\times$ for a minimal
resolution of 1\%.
}

\agpneutral{
Finally, we relate the runtimes of our algorithms to the
sample complexities with the scatter plot presented in
Figure~\ref{fig:sample-to-runtime}. The points on this plot represent
the number of samples versus the total execution times
of our sampling algorithms (excluding \scan) for varying dataset
sizes. As is evident, the runtime is directly proportional to the number
of samples.
% we can see, there is a clear relationship between
% the number of samples and the runtime of our algorithms.
%As we have shown, the runtime performance of our algorithms varies
%approximately linearly with respect to the sample complexity.
With this in mind,
for the rest of the synthetic datasets, we focus on sample complexity
because we believe it provides a more insightful view into the behavior
of our algorithms as parameters are varied. We return to
runtime performance when we consider real datasets in
Section~\ref{sec:real-exp}.
}

\stitle{Variation of Sampling and Accuracy with $\delta$:} 
We now measure how $\delta$ 
(the user-specified probability of error) affects the number of samples and accuracy.

\vspace{-7pt}
\begin{framed}
\vspace{-5pt}
\noindent {\em Summary:} For all algorithms, the percentage sampled
decreases as $\delta$ increases, but not by much.
The accuracy, on the other hand, stays  
constant at 100\%, independent of $\delta$. 
Sampling any less to estimate the same confidence intervals
leads to significant errors\techreport{, indicating that
the confidence intervals cannot shrink by much}.
\vspace{-5pt}
\end{framed}
\vspace{-7pt}

Figure~\ref{fig:mixture.mean.s} shows the effect of varying $\delta$
on the sample complexity for the six algorithms.
As can be seen in the figure, the percentage of data sampled 
reduces but does not go to $0$ as $\delta$ increases.
This is because the amount of
sampling (as in Equation~\ref{eq:samp-complexity}) 
is the sum of three quantities,
one that depends on $\log k$, the other on $\log \delta$,
and another on $\log \log (1/\eta_i)$. 
The first and last quantities are independent of $\delta$, 
and thus the number of samples required is non-zero 
even as $\delta$ gets close to $1$.
The fact that sampling is non-zero when $\delta$
is large is somewhat disconcerting; to explore whether
this level of sampling is necessary, and whether
we are being too conservative, we examine 
the impact of sampling less on accuracy (i.e., whether
the algorithm obeys the desired visual property).

% \srm{The follow is WAY too long for this result.  We should summarize the effects
% of the heuristic factor experiment in one paragraph.}

We focus on \ifocusr\
\techreport{(similar results are observed for 
all algorithms, and other distributions)}
and consider the impact of shrinking confidence intervals
at a rate faster than prescribed by \ifocus\ in Line 6 of
Algorithm~\ref{alg:pseudocode}.
We call this rate the {\em heuristic factor}:
a heuristic factor of 4 means that we divide the confidence
interval as estimated by Line 6 by 4, thereby ensuring 
that the confidence interval overlaps are fewer in number,
allowing the algorithms to terminate faster.
We plot the average accuracy (i.e., the fraction
of times the algorithm violates the visual ordering property)
as a function of the heuristic factor in Figure~\ref{fig:mixture.factor.a}
for $\delta = 0.05$ (other $\delta$s give identical figures,
as we will see below). 
\agpcomment{\alkim{This isn't quiet right. The heuristic factor of 4 just means we
  sample 4 times less, that doesn't translate to a 4x shrinkinage in
confidence interval.}}
% Note that the average accuracy
% stays the same independent of $\delta$, hence we have chosen 
% to show it for one $\delta$.

First, consider heuristic factor 1, which directly corresponds to
\ifocusr. As can be seen in the figure, 
\ifocusr\ has 100\% accuracy:
the reason is that \ifocusr\ ends up sampling a constant amount
to ensure that the confidence intervals do not overlap,
independent of $\delta$, enabling it to have perfect accuracy for
this $\delta$.
In fact, we find that 
all our 6 algorithms have accuracy 100\%,
independent of $\delta$ and the data distributions; 
thus, our algorithms not only provide much lower sample complexity,
but also respect the visual ordering property on all datasets.

Next, we see that as we increase the heuristic factor,
the accuracy immediately decreases (roughly monotonically) below 100\%, 
Surprisingly, even with a heuristic factor
of $2$, we start making mistakes at a rate 
greater than $2-3\%$ independent of $\delta$.
Thus, even though our sampling is conservative,
{\em we cannot do much better, and are likely to make errors} 
by shrinking confidence intervals
faster than prescribed by Algorithm~\ref{alg:pseudocode}.
To study this further, we plotted the same graph for the hard case 
with $\gamma = 0.1$ (recall that $\gamma = \eta$ for this case), 
in Figure~\ref{fig:hard.0.1.a}.
Here, once again, for heuristic factor 1, i.e., \ifocusr, 
the accuracy is 100\%. 
On the other hand, even with a heuristic factor of 1.01,
where we sample just 1\% less to estimate the same confidence interval,
the accuracy is {\em already less than 95\%}.
With a heuristic factor of 1.2, the accuracy is less than 70\%!
This result indicates that we cannot shrink our confidence intervals
any faster than \ifocusr\ does, since we may end up making up making 
far more mistakes than is desirable---even sampling just $1\%$ less
can lead to critical errors.

Overall, the results in Figures~\ref{fig:mixture.factor.a} and \ref{fig:hard.0.1.a}
are in line with our theoretical
lower bound for sampling complexity, which holds no matter
what the underlying data distribution is.
Furthermore, we find that algorithms backed
by theoretical guarantees are necessary 
to ensure correctness across all data distributions
(and heuristics may fail at a rate higher than $\delta$).

\stitle{Rate of Convergence:} 
In this experiment, we measure the rate of convergence of the \ifocus\ algorithms
in terms of the number of groups that still need to be sampled as the algorithms run.

\vspace{-7pt}
\begin{framed}
\vspace{-5pt}
\noindent {\em Summary:} Our algorithms converge
very quickly to a handful of active groups. Even when there
are still active groups, the number of incorrectly ordered groups is
very small\techreport{ --- thus, our algorithms can be used
to provide incrementally improving partial results}.
\vspace{-5pt}
\end{framed}
\vspace{-7pt}
Figure~\ref{fig:mixture.active_intervals.c} shows 
the average number of active groups 
as a function of the amount of sampling performed for \ifocus, over a
set of 100 datasets of size 10M.
It shows two scenarios:
$0$, when the number of samples is averaged across all 100 datasets and
$3M$, when we average across all datasets where at
least three million samples were taken.
For $0$, on average, the number of active groups after the first 
1M samples (i.e., 10\% of the 10M dataset),
is just 2 out of 10,
and then this number goes down slowly after that.
The reason for this is that, with high probability, there will be
two groups whose $\mu_i$ values are very close to each other. 
So, to verify if one is greater than the other, 
we need to do more sampling for those two groups, as compared to 
other groups whose $\eta_i$ (the distance to the closest
mean) is large---those
groups are not active beyond 1M samples.
For the $3M$ plot, we find that the number of
samples necessary to reach 2 active groups is larger,  
close to 3.5M for the 3M case.

Next, we investigate if the current estimates $\nu_i, \ldots, \nu_k$ respect the 
correct ordering property, even though some groups are still active.
To study this, we depict the number of incorrectly ordered pairs
as a function of the number of samples taken, once again for the two
scenarios described above.
As can be seen in Figure~\ref{fig:mixture.incorrect_pairs.c}, even though the number of active groups is
close to two or four at 1M samples, the number of incorrect pairs is very close
to 0, but often has small jumps --- indicating that the algorithm
is correct in being conservative and estimating that the we haven't yet identified
the actual ordering. 
In fact, the number of incorrect pairs
is non-zero up to as many as 3M samples, indicating that
we cannot be sure about the correct ordering without taking that
many samples.
At the same time, since the number of incorrect pairs is small,
if we are fine with displaying somewhat incorrect results,
we can show the current results to to the user.

\begin{figure*}[!t]
\centering
\subfigure{
	\includegraphics[height=1.5in]{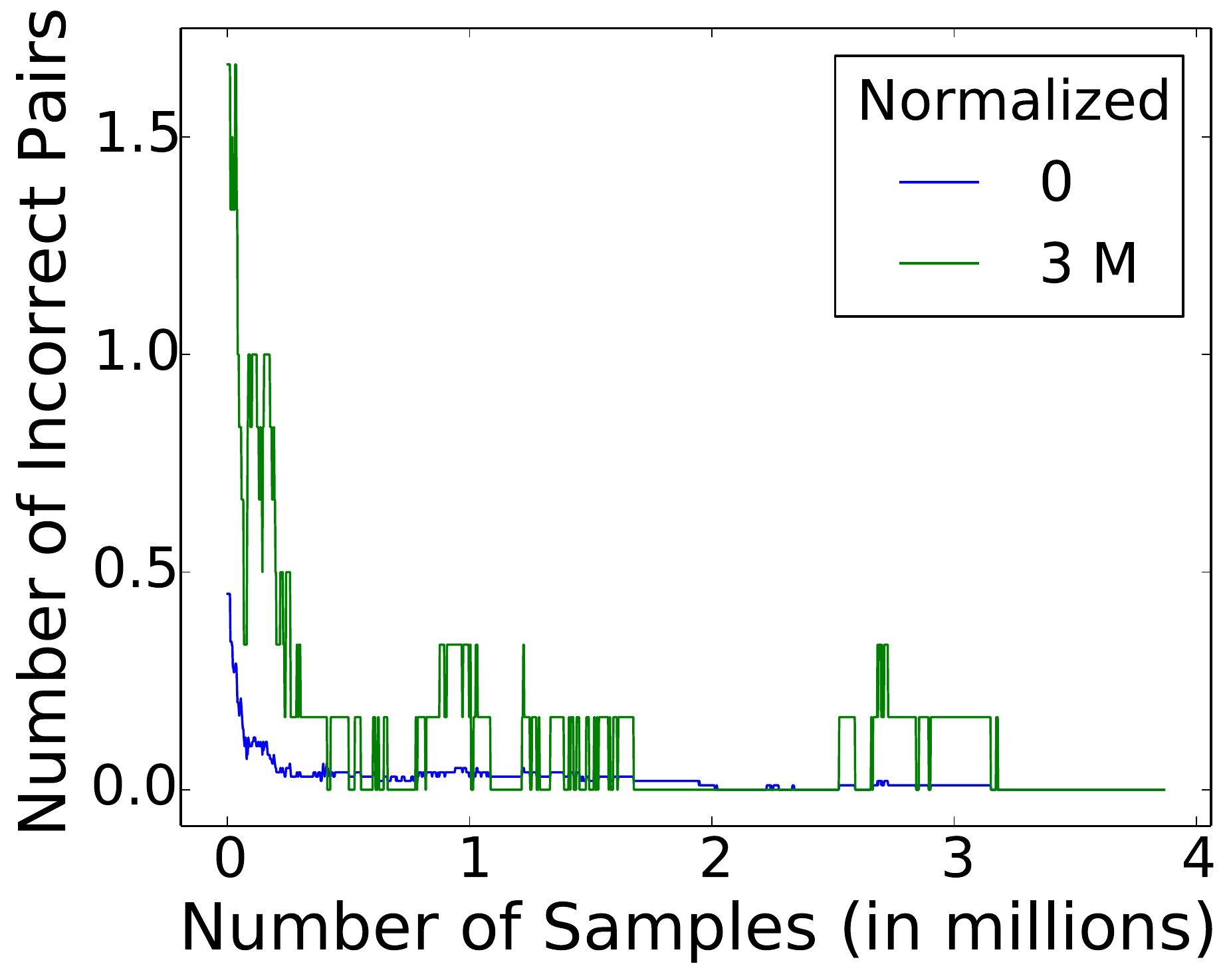}
	\label{fig:mixture.incorrect_pairs.c}
}
\subfigure{
	\includegraphics[height=1.5in]{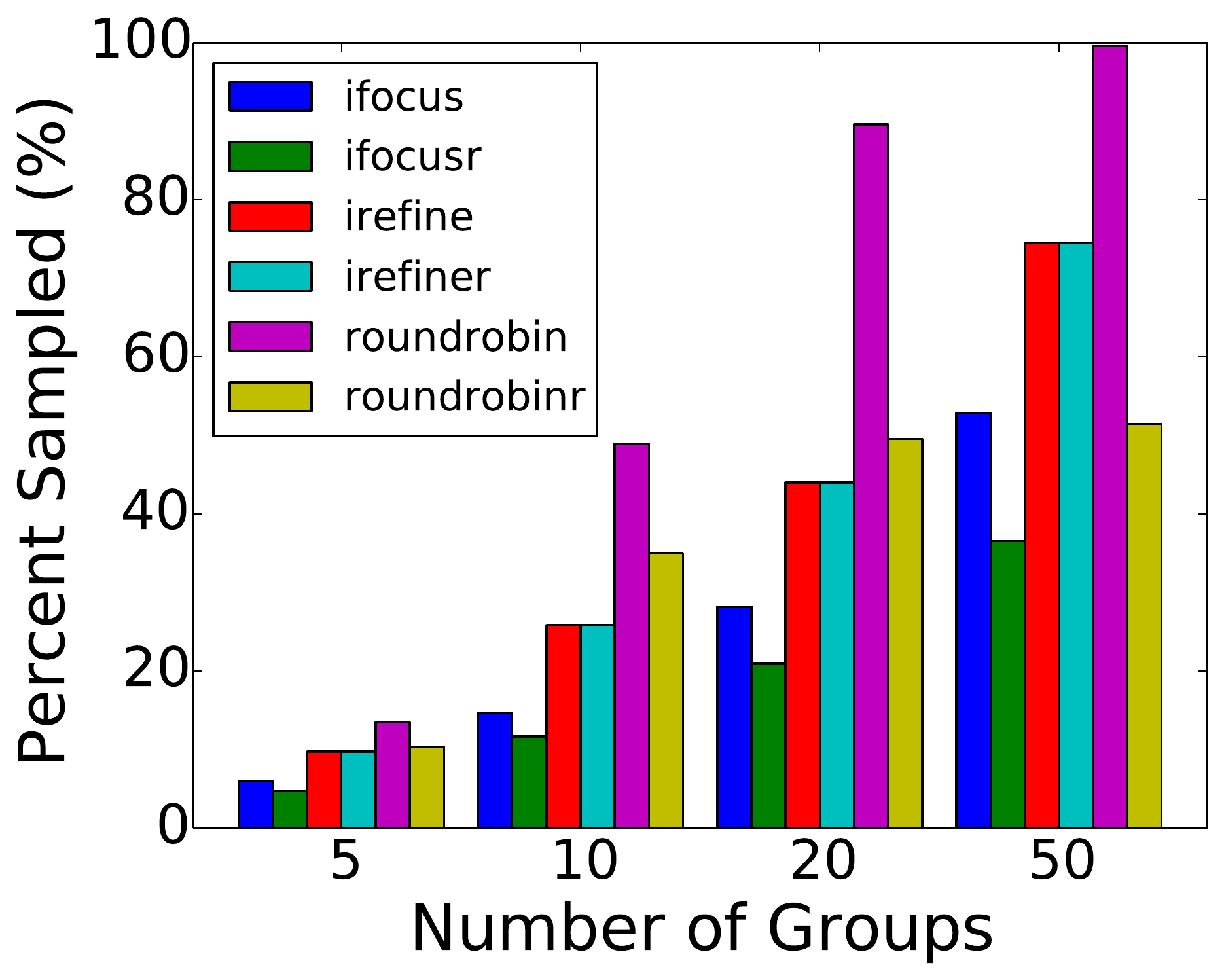}
	\label{fig:mixture.num_groups.mean.s}
}
 \subfigure{
 	\includegraphics[height=1.5in]{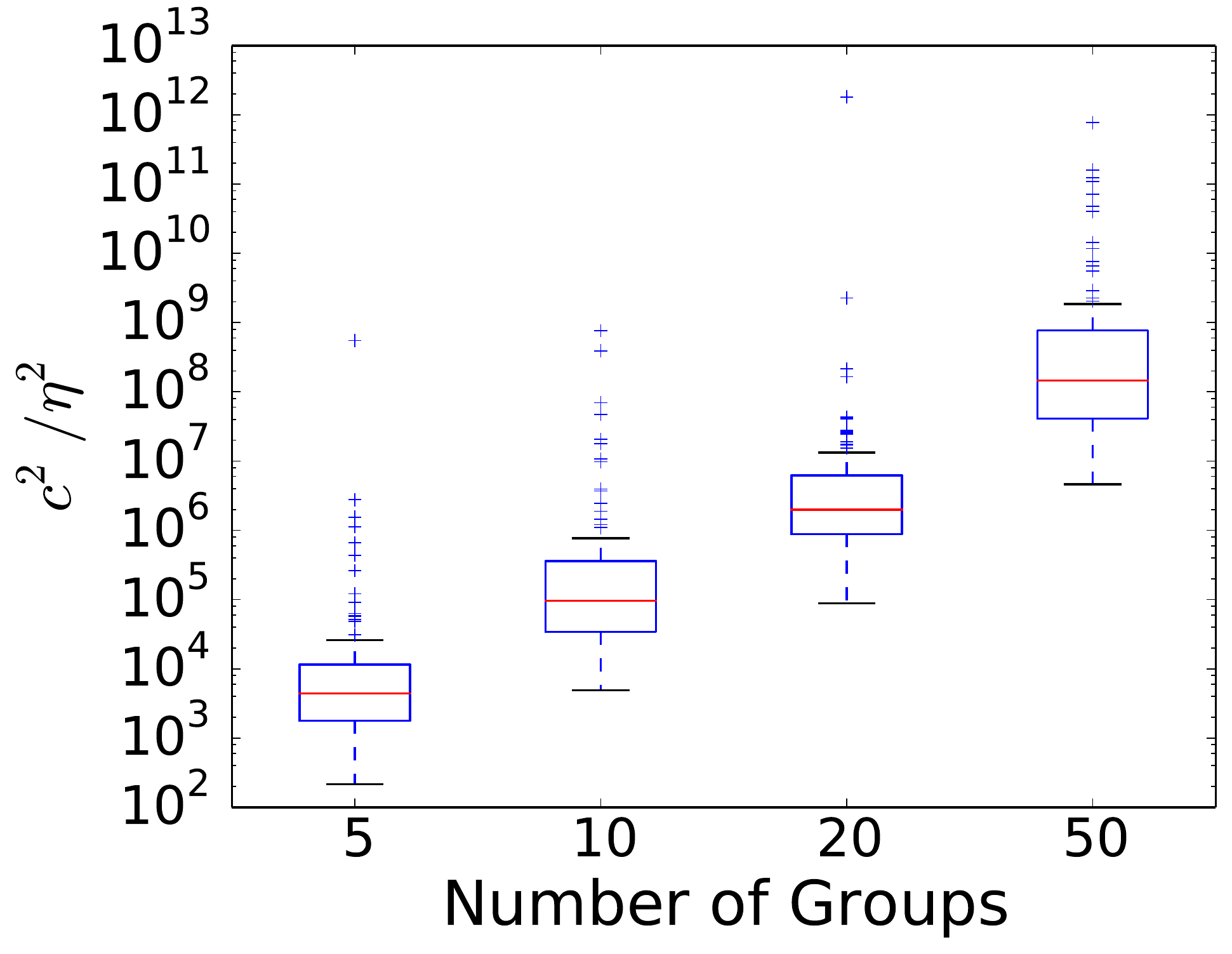}
 	\label{fig:mixture.num_groups.d}
 	%placeholder
 }
 \vspace{-10pt}
\caption{(a) Studying the number of incorrectly ordered pairs as computation proceeds (b) Impact of number of groups
on sampling (c) Evaluating the difficulty as a function of number of groups}
\vspace{-15pt}
\end{figure*}

\stitle{Variation of Sampling with Number of Groups:}
We now look at how the sample complexity varies with the number of groups.

\vspace{-7pt}
\begin{framed}
\vspace{-5pt}
\noindent {\em Summary:} As the number of groups increases, the amount of 
sampling increases for all algorithms
as an artifact of our data generation process\techreport{,
however, our algorithms continue to perform 
significantly better than other algorithms}.
\vspace{-5pt}
\end{framed}
\vspace{-7pt}

To study the impact of the number of groups on sample complexity,
we generate 100 synthetic datasets 
of type mixture where the number of groups varies from 5 to 50,
and plot the percentage of the dataset sampled 
as a function of the dataset size.
Each group has 1M items.
We plot the results in Figure~\ref{fig:mixture.num_groups.mean.s}.
As can be seen in the figure, our algorithms
continue to give significant gains even when the number
of groups increases from 5 to 50. 
However, we notice that the amount of sampling increases for 
\ifocusr\ as the number of groups is increased,
from less than 10\% for 5 groups to close to 40\% 
for 50 groups.

% \srm{I'm not sure it's important to show the following plot.  I suggest just saying ``The reason for this is that, as the number of groups grows,
% the likelihood that some means are very close also grows, which means that $\eta$ is smaller, and the problem
% is harder, requiring more samples.''}

The higher sample complexity 
can be attributed to the dataset generation process.
As a proxy for the ``difficulty'' of a dataset,
Figure~\ref{fig:mixture.num_groups.d} shows the average $c^2/\eta^2$ as a function of the number of groups (recall
that $\eta$ is the minimum distance between two means, $c$ is the range
of all possible values,
and that the sample complexity depends on $c^2/\eta^2$)
The figure is a a box-and-whiskers plot
with the y-axis on a log scale.
Note that the average difficulty
increases from $10^4$ for 5 to $10^8$ for 50--a 4 orders of magnitude increase!
Since we are generating means for each group
at random, it is not surprising that the
more groups, the higher the likelihood
that two randomly generated means will be close to each other. 

\papertext{
\stitle{Additional Experiments:}
In the extended technical report~\cite{tr}, we present additional experiments, 
including:
\begin{asparaitem}
\item {\em Dataset Skew:} Our algorithms continue to provide significant gains in the presence 
of skew in the underlying dataset.
\item {\em Variance:} Sample complexities of our algorithms vary slightly with variance;
sampling increases by 1-2\% as variance increases.
\end{asparaitem}
}

\techreport{
\stitle{Variation of Sampling with Proportion of Dataset:}
We now study the impact of skew on the algorithms.
\vspace{-7pt}
\begin{framed}
\vspace{-5pt}
\noindent {\em Summary:} Our algorithms continue to provide significant gains
in the presence of skew in the underlying dataset.
\vspace{-5pt}
\end{framed}
\vspace{-7pt}
We generated a variation of our 1M mixture dataset,
where we vary the fraction of the dataset that belongs
to the first group from 10\% to 90\%, while
the remaining fraction is equally distributed among the remaining
9 groups.
The results are depicted in Figure~\ref{fig:mixture.group0_prop.mean.s},
where we once again show the amount of sampling as a function of the proportion 
of the dataset that the first group occupies.
As can be seen in the figure, the relative gains of our \ifocus\ and \ifocusr\
algorithms continue to hold even when the proportion is 90\% (a highly skewed case).
Note also that the amount of sampling goes down as the proportion increases:
this is an artifact of our dataset generation mechanism.
Since the dataset is generated randomly, the odds that the first group is 
part of the set of active intervals stays fixed,
while if the first group was indeed among the active groups after
1M samples were taken, then we may need a lot more samples when the
first group has a larger number of tuples, as compared to the other groups.
Thus, the amount of total sampling goes down as the amount of skew increases.

\stitle{Variation of Sampling with Standard Deviation:}
We now examine how the variance of the data affects the number of samples.

% \srm{this result is boring -- could be briefly summarized in the previous section by saying ``In addition to looking at how the sample
% complexity varies with the number of groups, we looked at how changing the variance of the groups affects the results.  The results are similar to increasing the number of groups:  as variance increases, there is a greater probability that groups will overlap, making $\eta$ larger and the
% problem harder.''}

\vspace{-7pt}
\begin{framed}
\vspace{-5pt}
\noindent {\em Summary:} For truncnorm, as the
standard deviation increases, the amount of sampling
increases slightly.
\vspace{-5pt}
\end{framed}
\vspace{-7pt}

To study the impact of the standard deviation of the
groups in the dataset on the sampling performed,
we focus on the truncnorm distribution, wherein
each group is generated from a truncated
normal distribution with a fixed standard deviation.
We plot the average percentage of the dataset sampled
by \ifocusr\
as a function of the $\delta$ (desired accuracy),
for various values of the standard deviation of the
groups in the dataset.
The results are depicted in Figure~\ref{fig:truncnorm.std.ifocus.mean.s}.
As can be seen in the figure, the percentage 
sampled is higher for larger standard deviations,
but not by much (less than a 1-2\% change
across a range of standard deviations and $\delta$s).

To understand why the amount sampled for higher standard
deviations is higher, we plot the average 
$c^2/\eta^2$ as a function of 
the standard deviation as a box-and-whiskers plot
with the y-axis on a log scale in Figure~\ref{fig:truncnorm.std.ifocus.d}.
As can be seen in the figure, once again we find that
the datasets generated with a higher standard deviation
indeed have a higher ``difficulty'' (i.e., $c^2/\eta^2$),
and so therefore, require more samples.

\begin{figure*}[t]
\centering
\subfigure{
	\includegraphics[height=2.0in]{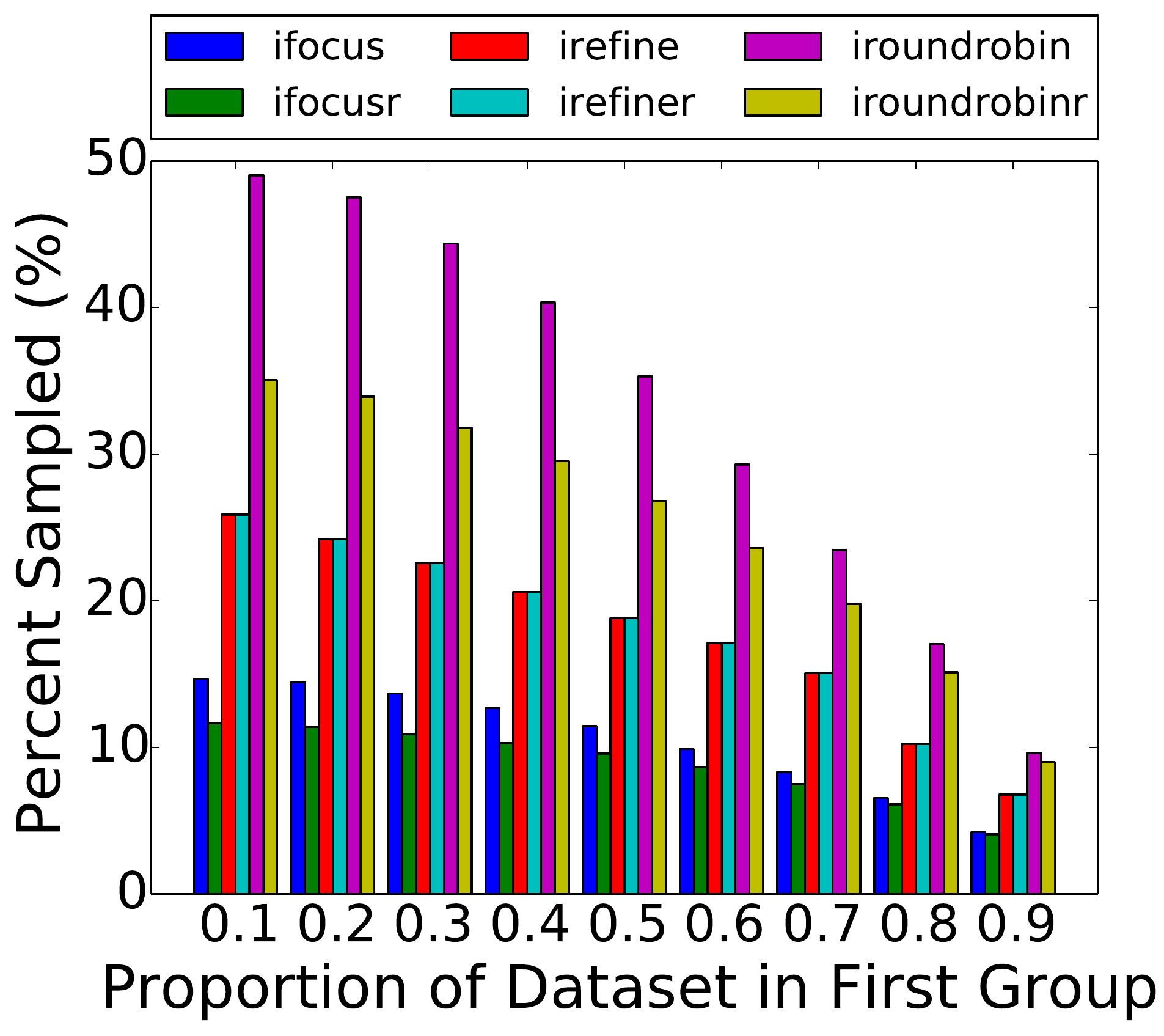}
	\label{fig:mixture.group0_prop.mean.s}
}
\hspace{-5pt}
\subfigure{
	\includegraphics[height=1.725in]{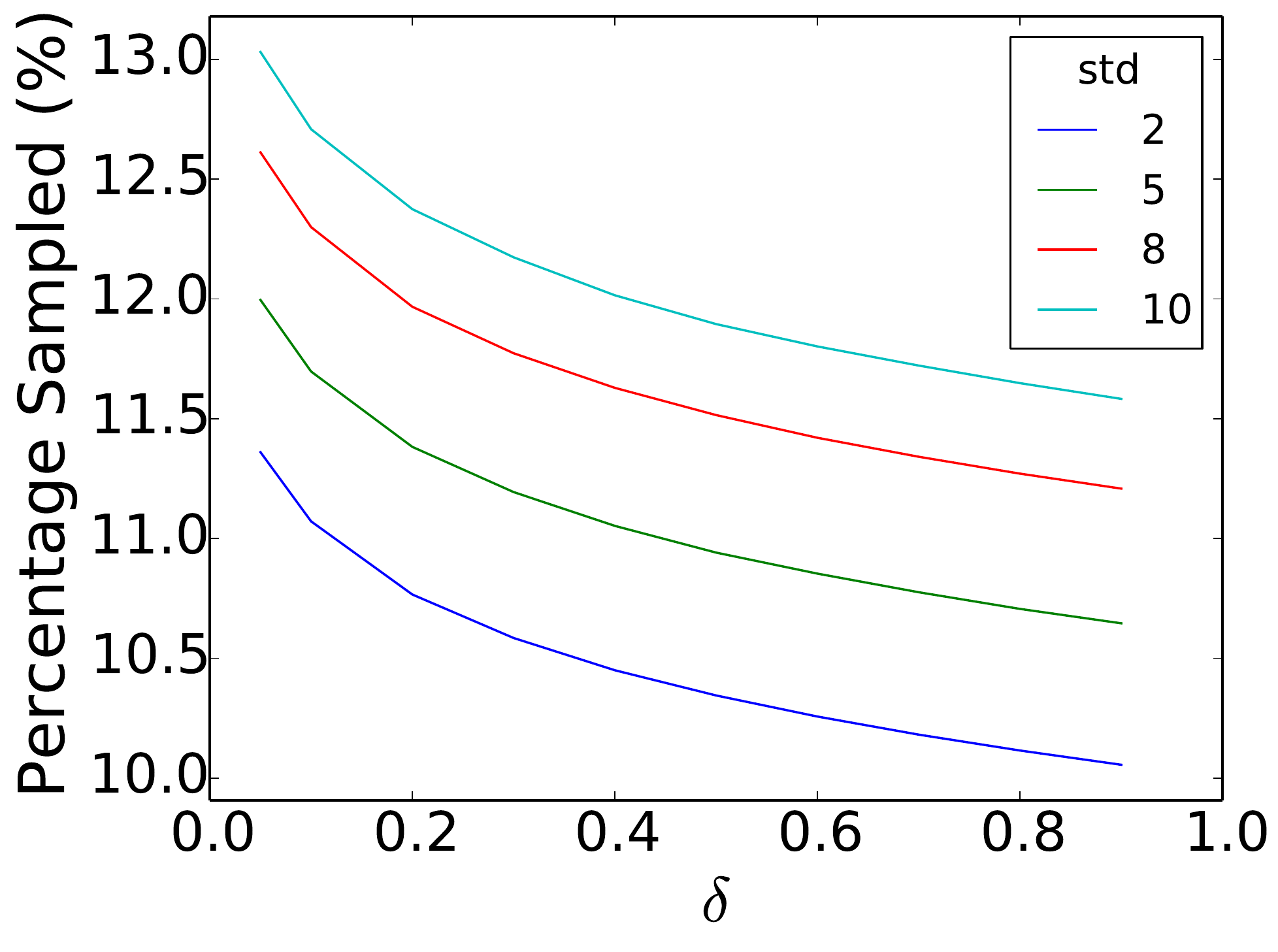}
	\label{fig:truncnorm.std.ifocus.mean.s}
}
\hspace{-5pt}
\subfigure{
	\includegraphics[height=1.725in]{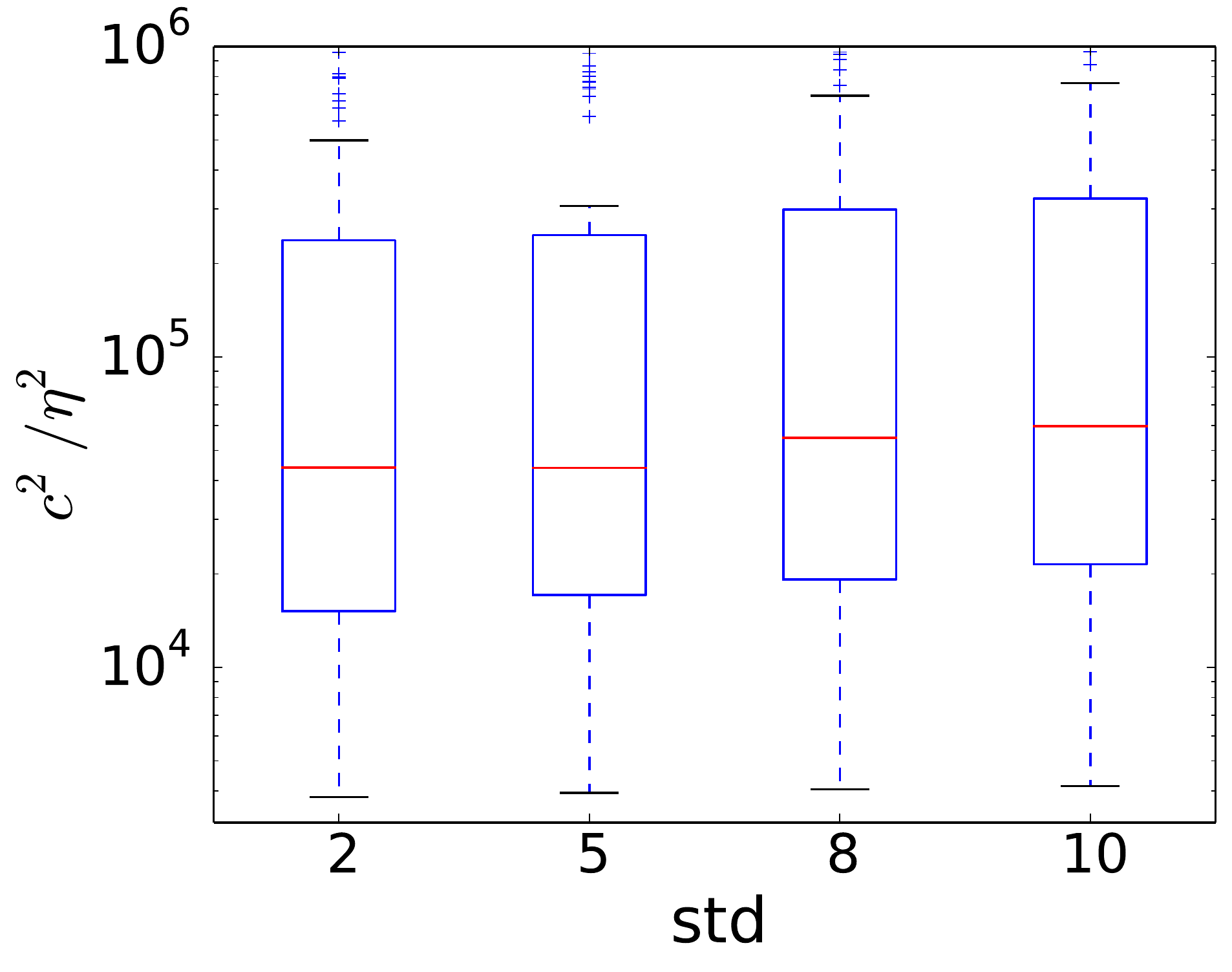}
	\label{fig:truncnorm.std.ifocus.d}
	%placeholder
}
\vspace{-5pt}
\caption{(a) Impact of the proportion of the first group on sampling (b) Impact of standard deviation on sampling  for truncnorm (c) Evaluating the difficulty on varying standard deviation for truncnorm}
\vspace{-15pt}
\end{figure*}
}%techreport

%!TEX root=main.tex

\subsection{Real Dataset Experiments}
We next study the impact of our techniques on a real dataset.

\vspace{-7pt}
\begin{framed}
\vspace{-5pt}
\noindent {\em Summary:} \ifocus\ and \ifocusr\ take 
 50\% fewer samples than
 \ir\ irrespective of the attribute
 visualized.
\vspace{-5pt}
\end{framed}
\vspace{-7pt}

For our experiments on real data, we used a flight records data set~\cite{flight-data}. 
The data set contains the details of all flights within the USA from
1987--2008, with nearly 120 million records, taking up 
12 GB uncompressed. 
From this flight data, we generated datasets of sizes 120 million
records (2.4GB) and scaled-up 1.2 billion (24GB) and 12 billion records (240GB) for our
experiments using probability density estimation.
%In addition to experimenting on the 120 million records, we scaled the
%dataset up to 1.2 billion and 12 billion records using density
%estimation to see how our algorithms would fare on these larger
%datasets.
We focused on comparing our best algorithms---\ifocus and
\ifocusr ($r$=1\%)---versus the conventional sampling---\ir. 
We evaluate the runtime performance for
visualizing the averages for three attributes: Elapsed Time,
Arrival Delay, and Departure Delay, grouped by Airline.
For all algorithms and attributes, the orderings returned were correct.

The results are presented in Table~\ref{tab:real-data}. 
The first four rows correspond to the attribute Elapsed Time. Here, \ir
takes 32.6 seconds to return a visualization, whereas \ifocus takes only
9.70 seconds (3$\times$ speedup) and \ifocusr takes only 5.04
seconds (6$\times$ speedup). We see similar speedups for Arrival Delay
and Departure Delay as well.
As we move from the $10^8$ dataset to $10^{10}$ dataset, we see the
run times roughly double for a 100$\times$ scale-up in the dataset.
The reason for any increase at all in the runtime comes from the
highly conflicting groups with means very close to one another. Our
sampling algorithms may read all records in the group for these groups
with with $\eta_i$ values. When the dataset size is increased to allow
for more records to sample from, our sampling algorithms take advantage
of this and sample more from the conflicting groups, leading to larger
run times.

\begin{table}
\centering
{\scriptsize
\begin{tabular}{|l|l|l|l|l|}\hline
Attribute & Algorithm & $10^8$ (s) & $10^{9}$ (s) & $10^{10}$
 (s) \\ \hline \hline
\multirow{3}{*}{Elapsed Time}
& \ir & 32.6 & 56.5 & 58.6 \\ \cline{2-5}
& \ifocus & 9.70 & 10.8 & 23.5 \\ \cline{2-5}
& \ifocusr(1\%) & 5.04 & 6.64 & 8.46 \\ \hline
\multirow{3}{*}{Arrival Delay}
& \ir & 47.1 & 74.1 & 77.5 \\ \cline{2-5}
& \ifocus & 29.2 & 48.7 & 67.5 \\ \cline{2-5}
& \ifocusr(1\%) & 9.81 & 15.3 & 16.1 \\ \hline
\multirow{3}{*}{Departure Delay}
& \ir & 41.1 & 72.7 & 76.6 \\ \cline{2-5}
& \ifocus & 14.3 & 27.5 & 44.3 \\ \cline{2-5}
& \ifocusr(1\%) & 9.19 & 15.7 & 16.0 \\ \hline
\end{tabular}
}
\vspace{-5pt}
\caption{Real Data Experiments}\label{tab:real-data}
\vspace{-18pt}
\end{table}

Regardless, we show that even on a real dataset, our sampling algorithms
are able to achieve up to a 6$\times$ speedup in runtime compared to
round-robin. We could achieve even higher speedups if we were willing to
tolerate a higher minimum resolution.

\label{sec:real-exp}

\techreport{%!TEX root=main.tex

\section{Extensions}\label{sec:extensions}
In this section, we describe a number of 
variations of our algorithm to handle
different scenarios: either stronger or weaker
conditions, or other types of queries or settings.

\subsection{Weaker Conditions}\label{sec:extensions-weaker}

We now describe extensions to \ifocus\ to handle
visual properties that are ``weaker'', i.e., 
require much less samples.

\subsubsection{Trends and Chloropleths} \label{sec:ext-trends}
When viewing trend lines instead of bar graphs---where
the $x$ axis is an ordinal attribute such as 
time---we instead want comparisons between
consecutive pairs of groups to be accurate
rather than all pairs.
We state the problem below:
\begin{problem}[AVG-Order-Trends]
Given a query $Q$, and values 
$c, \delta$, and an index on $X$, 
design a query processing algorithm 
returning estimates $\nu_1, \ldots, \nu_k$
for $\mu_1, \ldots, \mu_k$
which is as efficient as possible in terms of sample complexity ${\cal C}$,
such that with probability greater than $1 - \delta$,
the ordering of $\nu_1, \ldots, \nu_k$
with respect to $\mu_1, \ldots, \mu_k$
is correct, where correctness is now defined as the following:
\begin{quote}
for all $i \in 1 \ldots (k - 1)$, if $\mu_i < \mu_{i + 1}$ then
$\nu_i < \nu_{i + 1}$ if $\mu_i < \mu_{i + 1}$
and vice versa.
\end{quote}
\end{problem}
\stitle{Solution:}
The \ifocus\ algorithm generalizes easily
to the scenario where we only care about 
comparisons between neighboring groups instead of 
between all groups. 
In this case, we get the same sample complexity as now except that we 
replace the definition of $\eta_i$ to be $\eta_i^* = \min\{\tau_{i-1,i},\tau_{i,i+1}\}$,
and our definition of {\em active} now changes to all
groups whose confidence intervals overlap with confidence intervals
of neighboring groups (rather than all groups).

Similarly,
if, instead of a trend-line, we wished to generate a chloropleth
(i.e., heat map) where adjacent regions are correctly ordered with respect to 
each other (or, even if we wanted to ensure that 
the regions that are close by are correctly ordered with respect to each other),
then we can simply redefine {\em active} to mean all groups (here regions)
whose confidence intervals overlap with confidence intervals
of groups that are close by.

\subsubsection{Top-$t$ Results}
In many cases, the analyst is specifically interested in examining
the top-$t$ or bottom-$t$ groups rather than all the groups.
This is especially important if the number of groups is so large
that the analyst cannot easily look at them all at once.
In such a scenario, we need to make sure that 
we are confident about the fact that the $t$ groups we display
are indeed the top or bottom $t$, 
and that these $t$ groups are ordered correctly with respect to each other.

\begin{problem}[AVG-Order-Top-$t$]\label{prob:relative-order-top-t}
Given a query $Q$, and values 
$c, \delta, t$, a minimum resolution $r$, and an index on $X$, 
design a query processing algorithm 
returning estimates $\nu'_1, \ldots, \nu'_t$
for the largest $\mu'_1, \ldots, \mu'_t$
which is as efficient as possible in terms of sample complexity ${\cal C}$,
such that with probability greater than $1 - \delta$,
the ordering of $\nu_1', \ldots, \nu_t'$
with respect to $\mu_t', \ldots, \mu_t'$
is correct.
\end{problem}
\stitle{Solution:} For this variant, our definition of {\em active}
is now the groups for which either the confidence intervals
overlap other confidence intervals AND we are not yet sure if they 
are part of the top-$t$ or not.
As soon as we are sure that a group is not part of the top-$t$,
based on their confidence interval, we remove them from the
set of active groups.
This approach once again guarantees correctness with probability
greater than $1-\delta$.
\agpcomment{The sample complexity of this approach is the following:
\begin{align}\label{eq:samp-complexity}
O\left( c^2 \sum_{i=1}^k \frac{\log(\frac{k}{\delta}) + \log \log(\frac{1}{\eta_i})}{\eta_i^{\,2}}\right)
\end{align}}

\subsubsection{Allowing Mistakes} \label{sec:ext-weaker}
In order to generate our visualizations quickly, consider
the scenario when we are fine with making errors on $\gamma$\%
of the comparisons 
(in addition to $\delta$ probability error overall)
Thus, we may be able to eliminate wasting effort on
the most tricky comparisons, and focus instead on the
easy ones.
The problem is now:
\begin{problem}[AVG-Order-Mistakes]\label{prob:relative-order-mistakes}
Given a query $Q$, and values 
$c, \delta, \gamma$, and an index on $X$, 
design a query processing algorithm 
returning estimates $\nu_1, \ldots, \nu_k$
for $\mu_1, \ldots, \mu_k$
which is as efficient as possible in terms of sample complexity ${\cal C}$,
such that with probability greater than $1 - \delta$,
the ordering of $\nu_1, \ldots, \nu_k$
with respect to $\mu_1, \ldots, \mu_k$
is correct, where correctness is now defined as the following:
\begin{quote}
for $\gamma$ fraction or more of the pairs $(i, j), i \neq j$, the following holds:
if $\mu_i < \mu_j$ then $\nu_i < \nu_j$ and vice versa.
\end{quote}
\end{problem}
\stitle{Solution:} The algorithm for this problem is easy to state as a modification
of \ifocus: we simply keep track of the fraction of 
correct pairs that we correctly know the ordering of
(these are simply pairs of all inactive groups).
Once the desired fraction $\gamma$ is met, the
algorithm terminates and the estimates are returned.

% \subsubsection{Dropping Low Cardinality Bars}
% \agp{drop..}

\subsection{Stronger Conditions}\label{sec:extensions-stronger}
We now describe extensions to \ifocus\ that are ``stronger'',
i.e., require more samples than \ifocus.

\subsubsection{Approximate Actual Values}\label{sec:ext-approx-actual}
We now consider the generalization where
in addition to providing a ordering guarantee,
we would like to ensure
that the estimated averages per group
is close to the actual averages. 
The problem, therefore, is:
\begin{problem}[AVG-Order-Actual]
Given a query $Q$, and values 
$c, \delta$, a minimum approximation value $d$, and an index on $X$, 
design a query processing algorithm 
returning estimates $\nu_1, \ldots, \nu_k$
for $\mu_1, \ldots, \mu_k$
which is as efficient as possible in terms of sample complexity ${\cal C}$,
such that with probability greater than $1 - \delta$,
the ordering of $\nu_1, \ldots, \nu_k$
with respect to $\mu_1, \ldots, \mu_k$
is correct, 
and for all $i \in 1 \ldots k, |\nu_i - \mu_i| \leq d$.
\end{problem}
\stitle{Solution:}
For this variant, 
we first ensure a minimum amount of sampling $m$
such that $\varepsilon \geq d/2$. 
Once we perform the minimum amount of sampling,
the \ifocus\ algorithm proceeds as before.
The sample complexity for this algorithm is 
the same as that for \ifocus\ (as listed in Theorem~\ref{thm:analysis}) 
except that $\eta_i$ is replaced with $\min\{\eta_i, d/2\}$.

\subsubsection{Partial Results} \label{sec:ext-partial}
Our \ifocus\ algorithm is provably optimal in that it
lets us get to the correct ordering with the least amount of samples.
In many cases, it would be useful to show the analyst the 
estimated averages of the 
groups whose values we are already confident about,
so that they can start viewing and analyzing ``partial'' results.

% Do our algorithms translate to provide the best partial
% results that improve over time? That is, can we show that at
% any point, our algorithm is doing the best it can in resolving the 
% groups (this is a stronger condition of optimality)?
% Perhaps slightly weaker is: how soon can the algorithm
% ``decide'' on a few groups so that it can display those
% groups quickly?

\begin{problem}[AVG-Order-Partial]\label{prob:relative-order-partial}
Given a query $Q$, and values 
$c, \delta$, a minimum resolution $r$, and an index on $X$, 
design a query processing algorithm 
returning estimates $\nu_1, \ldots, \nu_k$
for $\mu_1, \ldots, \mu_k$
which is as efficient as possible in terms of sample complexity ${\cal C}$,
such that with probability greater than $1 - \delta$,
the ordering of $\nu_1, \ldots, \nu_k$
with respect to $\mu_1, \ldots, \mu_k$
is correct; 
and the algorithm outputs each estimate $\nu_i$ 
as soon as it is sure about it.
\end{problem}
\stitle{Solution:} Our solution for this variant
is straightforward: we simply output the estimates
for each group as soon as they become inactive. 
We have the following guarantee: with probability $1-\delta$,
the ordering between all groups whose estimates
are output at any stage in the algorithm is correct.
Further, the sample complexity to get to the first $k'$ groups
being output is the following:
$$O\left( \frac{c^2 k'}{\eta_{k'}'} (\log(\frac{k}{\delta}) + \log \log(\frac{1}{\eta_{k'}'}))\right)$$
where $\eta_{k'}'$ is the smallest $\eta_i$ among the first
$k'$ groups to become inactive.

% At every iteration of the algorithm, we 
% do have a useful guarantee. Namely, when we are in the iteration for a given $\varepsilon$,
% we know that we can always display our estimates for the average $\mu_i$ for every group
% $i$ that satisfies $\eta_i > 2\varepsilon$. (So, informally, the groups whose averages 
% are ``far'' from every other average can be displayed very quickly.)

% \subsubsection{Order Of Magnitude Considerations}\label{sec:order}
% \agp{drop if no results..}
% Can we enforce somewhat stronger considerations than
% comparisons? For instance, if an aggregate for a group
% is actually $k$-times the aggregate for another group 
% in reality, can we ensure that it is within 
% $\alpha k$ and $\frac{k}{\alpha}$ in the visualization 
% that is displayed?

\subsection{Other Settings and Queries}\label{sec:extensions-complex}
Our techniques can be adapted to a number of other 
settings or more complex queries.

\subsubsection{Different Aggregation Functions: SUM}
So far, we have focused our attention on estimating
the ${\tt AVG}$ value of each group, ensuring that
the correct ordering property is maintained.
We now discuss extending our algorithms to estimate
${\tt SUM}$ instead of average.

\stitle{Known Group Sizes:} If we know the number of elements in each group,
the two problems are equivalent;
the sum $\sigma_i$ of the elements in the group $S_i$ 
is related to the average $\mu_i$ value
of the elements in the group via the basic identity 
$\sigma_i = \mu_i \cdot |S_i|$.
However, the algorithm needs to change slightly in order to 
correctly compute confidence intervals in this setting.
Specifically, in Statement 6 and 9, we multiply the right hand side
with $|S_i|$;
and our estimates $\nu_i$ now correspond to
estimates of $\sigma_i$ instead of $\mu_i$.
\papertext{The algorithm is listed in the extended
technical report~\cite{tr}.}

\techreport{
The algorithm is listed in Algorithm~\ref{alg:sum-known-pseudocode}.
\begin{algorithm}[h!]
\KwData{$S_1,\ldots,S_k,\delta$}
Initialize $m \gets 1$\;
Draw $m$ samples from each of $S_1,\ldots,S_k$ to provide initial estimates $\nu_1,\ldots,\nu_k$\;  
Initialize $A = \{1, \ldots, k\}$\;
\While{$A \neq \emptyset$}{
  $m \gets m+1$\;
  \For{each $i \in A$}{
    $\varepsilon_i = c \, |S_i| \, \sqrt{\frac{2\log\log(m) + \log(\pi^2k/3\delta)}{2m}}$\;
  }
  \For{each $i \in A$}{
    Draw a sample $x$ from $S_i$\;
    $\nu_i \gets |S_i| (\frac{m-1}{m} \nu_i + \frac1{m}x$)\;
    }
  \For{each $i \in A$}{
    \If{$[\nu_i - \varepsilon_i, \nu_i + \varepsilon_i] \cap 
          \big(\bigcup_{j \in A \setminus \{i\}} [\nu_j - \varepsilon_j, \nu_j + \varepsilon_j]\big) = \emptyset$}{$A \gets A \setminus \{i\}$}
    } 
}
Return $\nu_1,\ldots,\nu_k$\;
\caption{\ifocus--Sum1}\label{alg:sum-known-pseudocode}
\end{algorithm}
}

\stitle{Unknown Group Sizes:}
When we don't know the number of tuples in each group, then our problem
becomes a bit more complicated, since we need to simultaneously estimate
both the sizes of the groups, as well as the average, in order
to be able to estimate ${\tt SUM}$ overall across all groups.
For the purposes of this discussion, we assume that we know
the total number of elements in all the groups; however,
our algorithm does not depend on this knowledge. 
If the total number of elements across all groups is known,
we can start reasoning about fractional sizes.
We let $s_i = \frac{|S_i|}{\sum_{j=1}^k |S_j|}$ 
to denote the fractional size of $S_i$. Then
$\sigma_i = s_i \mu_i$ is the normalized sum of the elements in the set $S_i$. The problem
of estimating the sums ensuring correct ordering 
is identical to that of estimating the 
normalized sums with correct ordering, so we now focus on the latter.

Using our \ntail\ indexes, when we retrieve an additional tuple from $S_i$, 
we can also estimate the number of tuples we needed to skip over until
we reach the tuple that belongs to $S_i$.
\ntail's in-memory bitmap indexes allow us to retrieve
this information without doing any disk seeks.
This number allows us to get unbiased estimates for the
normalized sums $s_1 \mu_1,\ldots,s_k \mu_k$. 
Then if $x$ is a random element from $S_i$ and $z$ is a random unbiased estimate of $s_i$, 
the random variable $x\cdot z$ is an unbiased estimate of $\sigma_i$. The random variable
$x \cdot z$ is also in the range $[0,c]$, so we again can apply Hoeffding inequalities
to derive and analyze an algorithm that is very similar to the original \ifocus.
The only difference between \ifocus\ and this new algorithm
is that we simultaneously get samples for $x$ and $z$;
the confidence interval computation stays unchanged.

The fact that the confidence interval computation looks 
exactly the same as when we're computing the
average is somewhat surprising,
since we're trying to estimate the size of each group
as well as the average---one may expect 
the confidence intervals to be larger than before.
In the worst case, when all the groups have the same size
(which is identical to when we're estimating the average),
this is essentially what happens.
If the group $S_i$ has average value $\mu_i$, then its normalized sum
is now $\mu_i/k$, i.e., 
the normalized sums are all $k$ times smaller than the corresponding averages.
But since our confidence intervals 
shrink at the same pace in the average and normalized sum cases, 
it will take much longer to be small enough to avoid overlaps for normalized sums. 
Specifically, this will require a number of samples that is roughly $k^2$ times 
larger in the normalized sum case than in the average case.
However, we do expect groups to be widely varying in size,
in which case, we do not expect the additional $k^2$ factor 
to affect the sample complexity.

\papertext{The pseudocode for the Algorithm can be found
in the extended technical report~\cite{tr}.}
\techreport{The pseudocode for the Algorithm can be found 
in Algorithm~\ref{alg:pseudocode-sum-complete}.
\begin{algorithm}[h!]
\KwData{$S_1,\ldots,S_k,\delta$}
Initialize $m \gets 1$\;
Draw $m$ samples from each of $S_1,\ldots,S_k$ to provide initial estimates $\nu_1,\ldots,\nu_k$\;  
Initialize $A = \{1, \ldots, k\}$\;
\While{$A \neq \emptyset$}{
  $m \gets m+1$\;
  $\varepsilon_i = c \, \sqrt{\frac{2\log\log(m) + \log(\pi^2k/3\delta)}{2m}}$\;
  \For{each $i \in A$}{
    Draw a sample $x$ from $S_i$\;
    Draw an estimate $z$ of the size $s_i$\;
    $\nu_i \gets \frac{m-1}{m} \nu_i + \frac1{m}xz$\;
    }
  \For{each $i \in A$}{
    \If{$[\nu_i - \varepsilon_i, \nu_i + \varepsilon_i] \cap 
          \big(\bigcup_{j \in A \setminus \{i\}} [\nu_j - \varepsilon_j, \nu_j + \varepsilon_j]\big) = \emptyset$}{$A \gets A \setminus \{i\}$}
    } 
}
Return $\nu_1,\ldots,\nu_k$\;
\caption{\ifocus--Sum2}\label{alg:pseudocode-sum-complete}
\end{algorithm}
}

\subsubsection{Different Aggregation Functions: COUNT}\label{sec:ext-diff-agg}
Naturally, estimating ${\tt COUNT}$ per group
is trivial if the number of
tuples per group is known. 
If the number of tuples is not known
(while the total number of tuples is known),
we can simply apply the algorithm for 
${\tt SUM}$, while only getting samples for $s_i$, rather 
than $\sigma_i$.
Note that $s_i \in [0,1]$ rather than $[0,c]$ like in
the previous case.
\agpcomment{VAR, MIN/MAX, not clear}

\subsubsection{Selection Predicates}\label{sec:ext-sel}
Consider the scenario when we have a query of the form:
\begin{quote}
\hspace{-15pt}${\tt SELECT}  \ \ X, \ \ {\tt AVG}(Y) \ \  {\tt FROM} \ \ R(X, \  Y, \ldots) \ \ {\tt GROUP} \ \ {\tt BY} \ \ X \ \ {\tt WHERE} \ \ $Pred 
\end{quote}
Here, we may have additional predicates on $X, Y$
or other attributes.
For instance, we may want to view the average delay of all airlines
whose delay is more than half an hour, 
i.e., the flights with a long delay.

\stitle{Solution:}
Our algorithms still continue to work 
even if we have selection predicates on 
one more more attributes,
as long as we have an index on the group-by attribute 
(the case where we do not have an index on the group-by 
attribute is captured in Section~\ref{sec:ext-no-indexes}).
Here, \ntail's bitmap indexes allow us 
to retrieve, on demand, tuples
that are from any specific group $S_i$,
and also satisfy the selection conditions specified.

\subsubsection{Multiple Group-bys}\label{sec:ext-group}
If $Q$ consists of multiple group bys, for instance:
\begin{quote}
\hspace{-15pt}${\tt SELECT}  \ \ X, \ Z, \ \ {\tt AVG}(Y) \ \  {\tt FROM} \ \ R(X, \  Y, \  Z) \ \ {\tt GROUP} \ \ {\tt BY} \ \ X, \ Z $
\end{quote}
Here, we may want to view a three-dimensional visualization
or a two-dimensional visualization with 
the cross-product of the $X, Z$ values on the X axis.
For instance, we may want to see the average
delay of all flights by airline and origin airport.

\stitle{Solution:} In this case, we can 
simply apply the same algorithms 
as long as we have either a joint index on $X$ and $Z$, 
or an index on either $X$ or $Z$.
When we have an index on just $X$ or just $Z$, 
the algorithms can still operate correctly, but
may have a higher sample complexity.
Say we have an index only on $X$. 
In such a case, we continue taking samples
from the groups with $X = x_i$,
as long as there is some value $z_j$,
such that the group corresponding to $(x_i, z_j)$
is still active.

\subsubsection{Multiple Aggregates Visualized}\label{sec:ext-mult-agg}
Consider the scenario when the analyst
wishes to visualize multiple aggregates, 
as the following query indicates:
\begin{quote}
\hspace{-15pt}${\tt SELECT}  \ \ X, \ {\tt AVG}(Z), \ {\tt AVG}(Y) \ \  {\tt FROM} \ \ R(X, \ Y,  \ Z) \ \ {\tt GROUP} \ \ {\tt BY} \ \ X $
\end{quote}

\begin{problem}[AVG-AVG-Order]\label{prob:relative-order-twice}
Given a query $Q$, and values 
$c, \delta$, a minimum resolution $r$, and an index on $X$, 
design a query processing algorithm 
returning estimates $\nu_{11}, \ldots, \nu_{1k}$
for $\mu_{11}, \ldots, \mu_{1k}$ (true averages of $Y$), 
and  $\nu_{21},$ $\ldots, $ $\nu_{2k}$
for $\mu_{21},$ $ \ldots, $ $\mu_{2k}$ (true averages of $Z$),
which is as efficient as possible in terms of sample complexity ${\cal C}$,
such that with probability greater than $1 - \delta$,
the ordering of $\nu_{i1}, \ldots,$ $ \nu_{ik}$
with respect to $\mu_{i1}, \ldots, \mu_{ik}$ 
is correct for both $i = 1, 2$.
\end{problem}
\stitle{Solution:} In this case, we apply \ifocus\
to the problem of ${\tt AVG}(Y)$ first (with $\delta$ set as $\delta/2$), 
while also simultaneously estimating ${\tt AVG}(Z)$.
Then, once there are no longer any more active groups, we
apply \ifocus\ to ${\tt AVG}(Z)$ (with $\delta$ set as $\delta/2$), 
starting at the estimates we already have. 
In the worst case, the sample complexity will be the
sum of the sample complexities of running the two independently,
but since the second iteration of \ifocus\ will start 
having sampled quite a few values per group for ${\tt AVG}(Z)$,
the samples taken for the second iteration of \ifocus\ is
likely to be much smaller than the first.

% \subsubsection{Unknown Number of Groups}\label{sec:ext-unknown}
% In the case where the number of groups is unknown
% (possibly due to a selection condition),
% we can first probe the index to check each group once
% and then focus on the non-empty groups
% \agp{hard to justify without \nt}

\subsubsection{No Indexes}\label{sec:ext-no-indexes}
We now consider the scenario when 
there is no index on the group-by attribute $X$.
The new problem in this scenario can be stated as the following:

\begin{problem}[AVG-Order-NoIndex]\label{prob:relative-order-noindex}
Given a query $Q$, and values 
$c, \delta$, design a query processing algorithm 
returning estimates $\nu_1, \ldots, \nu_k$
for $\mu_1, \ldots, \mu_k$
which is as efficient as possible in terms of sample complexity ${\cal C}$,
such that with probability greater than $1 - \delta$,
the ordering of $\nu_1, \ldots, \nu_k$
with respect to $\mu_1, \ldots, \mu_k$
is correct.
\end{problem}
We assume that no other indexes are present.
Without an index on $X$, 
we cannot sample from specific groups; we can only
get a random sample from any one of the groups.
However, we can still use Hoeffding's inequality
to decide when to terminate taking
random samples. 
If the number of tuples per group is roughly the same,
the performance of this algorithm 
would be similar to the performance of a round robin
approach that 
takes a sample from all groups no matter if they are active or not.
This approach, although poor compared to \ifocus, 
allows us to get away by sampling much less 
of the dataset (as we will see in Section~\ref{sec:exp}).

}

%!TEX root=main.tex

\section{Related Work}\label{sec:related}
The work related to our paper can be placed in a few categories:

\stitle{Approximate Query Processing:}
\agpneutral{There are two categories of related work in approximate query processing:
online, and offline. We focus on online first since it is more closely related to
our work.}

\agpneutral{Online aggregation~\cite{online-aggregation} is perhaps the most
related online approximate query processing work. It uses conventional
round-robin stratified sampling~\cite{casella} (like \ir)
to construct confidence intervals for estimates of averages of groups.
In addition, online aggregation provides an interactive tool that allows
users to stop processing of certain groups when their confidence is ``good enough''.
Thus, the onus is on the user to decide when to stop processing groups
(if not, stratified sampling is employed for all groups).
Here, since our target is a visualization with correct properties, 
\ifocus automatically decides when to stop processing groups.
Hence, we remove the burden on the user, and prevent the user
from stopping a group too early (making a mistake), 
or too late (doing extra work).}

\agpneutral{There are other papers that also use round-robin stratified sampling
for various purposes, primarily for COUNT estimation
respecting real-time 
constraints~\cite{hou89}, 
respecting accuracy constraints (e.g., 
ensuring that confidence intervals shrink to a pre-specified size) without indexes~\cite{hou88}, 
and with indexes~\cite{lnss93, hnss96}.}

\agpneutral{Since visual guarantees in the form of relative ordering
is very different from the kind of objectives prior work in
online approximate query processing considered, our techniques are quite different.
Most papers on online sampling for query processing, including \cite{hou88,hou89,lnss93,online-aggregation}, either use uniform random sampling or round-robin stratified sampling. Uniform random sampling is strictly worse than round-robin stratified sampling (e.g., if the dataset is skewed) and in the best case is going to be only as good, which is why we chose not to compare it in the paper.
On the other hand, we demonstrate that conventional sampling schemes like round-robin stratified sampling
sample a lot more than our techniques.}

\agpneutral{Next, we consider offline approximate query processing.  Over the
  past decade, there has been a lot of work on this topic; as
  examples,
  see~\cite{DBLP:conf/vldb/Gibbons01,dbo,wavelets}.
  Garofalakis et al.~\cite{Garofalakis:2001:AQP:645927.672356}
  provides a good survey of the area; systems that support offline
  approximate query processing include
  BlinkDB\cite{DBLP:conf/eurosys/AgarwalMPMMS13} and
  Aqua~\cite{Acharya:1999:AAQ:304182.304581}.  Typically, offline schemes
  achieve a user-specified level of accuracy by running the query on a
  sample of a database.  These samples are chosen a-priori, 
  typically tailored to a workload or a small set of queries~\cite{Acharya:2000:CSA:342009.335450,
    Ioannidis:1999:HAS:645925.671527,
    Babcock:2003:DSS:872757.872822,DBLP:conf/stoc/AlonMS96,914867}.  In our
  case, we do not assume the presence of a precomputed sample, since
  we are targeting ad-hoc visualizations.  Even when precomputing
  samples, a common strategy is to use Neyman
  Allocation~\cite{cochran2007sampling}, like in
  \cite{Chaudhuri:2007:OSS:1242524.1242526,joshi2008robust}, by
  picking the number of samples per strata to be such that the variance
  of the estimate from each strata is the same.  In our case, since we
  do not know the variance up front from each strata (or group), this
  defaults once again to round-robin stratified sampling. Thus, we
  believe that round-robin stratified sampling is an appropriate and
  competitive baseline, even here.}

\stitle{Statistical Tests:} There are a number of statistical tests~\cite{all-of-statistics, casella}
used to tell if two distributions are significantly different,
or whether one hypothesis is better than a set of hypotheses
(i.e., statistical hypothesis testing).
Hypothesis testing allows us to 
determine, given the data collected so far, whether we can reject
the null hypothesis.
The $t$-test~\cite{casella} specifically allows us to determine if 
two normal distributions are different from each other,
while the Whitney-Mann-U-test~\cite{mann-whitney} allows us to determine if
two arbitrary distributions are different from each other, 
None of these tests can be directly applied to 
decide where to sample from a collection of sets to
ensure that the visual ordering property is preserved.

\techreport{
\stitle{Noisy Sorting:}
Our work is also related to sorting with noisy comparisons, both in the context
of error-prone processing units~\cite{feige},  
or human workers~\cite{so-who-won, top-k}.
In our case, we cannot directly ask for comparisons between two groups, 
we can only get additional samples per group, and these additional samples
can help all the comparisons that the specific group is involved in.
}

\stitle{Visualization Tools:} 
Over the past few years, the visualization community has introduced a number of interactive visualization tools such as ShowMe, Polaris, Tableau, and Profiler~\cite{DBLP:journals/cacm/StolteTH08,DBLP:conf/avi/KandelPPHH12,DBLP:conf/sigmod/Hanrahan12}.  Similar visualization tools have also been introduced by the database community, including Fusion Tables~\cite{DBLP:conf/sigmod/GonzalezHJLMSSG10}, VizDeck~\cite{Key:2012:VSD:2213836.2213931}, and Devise~\cite{DBLP:conf/sigmod/LivnyRBCDLMW97}.
A recent vision paper~\cite{seedb-tr} has proposed a tool for recommending interesting visualizations of query results to users.
All these tools could benefit from the algorithms outlined in this paper to improve performance
while preserving visual properties.

\stitle{Scalable Visualization:}
\agpneutral{There has been some recent work on scalable visualizations from the information visualization community as well.
Immens~\cite{2013-immens} and Profiler~\cite{DBLP:conf/avi/KandelPPHH12} maintain a data cube in memory and use it
to support rapid user interactions. While this approach is possible when the dimensionality and cardinality is small (e.g., for simple map visualizations of a single attribute),
it cannot be used when ad-hoc queries are posed. A related approach uses precomputed image tiles for geographic
visualization~\cite{hotmap}.
}

\agpneutral{Other recent work has addressed other aspects of visualization scalability,
including prefetching and caching~\cite{doshi2003prefetching}, data reduction~\cite{burtini2013time} leveraging time series data mining~\cite{esling2012time}, clustering and sorting~\cite{guo2003coordinating,seo2005rank}, and dimension reduction~\cite{Yang:2003:VHD:769922.769924}. These techniques are orthogonal to our work, which focuses on speeding up the computation of a single visualization online.}

\agpneutral{Recent work from the visualization community has also demonstrated via user studies on simulations that users are satisfied with uncertain
visualizations generated for algorithms like online aggregation, as long as the visualization shows error bars~\cite{DBLP:conf/ldav/Fisher11,DBLP:conf/chi/FisherPDs12}.
This work supports our core premise, that analysts are willing to use inaccurate visualizations as long
as the trends and comparisons of the output visualizations are accurate.}

\stitle{Learning to Rank:} The goal of learning to rank~\cite{learning-to-rank} is the following:
given training examples that are ranked pairs of entities (with their features),
 learn a function that correctly orders these entities.
While the goal of ordering is similar, in our scenario we assume
no relationships between the groups, nor the presence of features
that would allow us to leverage learning to rank techniques.

%!TEX root=main.tex

\section{Conclusions}\label{sec:conc}

Our experience speaking with data analysts is indeed that they
prefer quick visualizations that look similar to visualizations
that are computed on the entire database. 
Overall, increasing interactivity
(by speeding up the processing of each visualization, even if it is approximate) 
can be a major productivity boost.
As we demonstrated in this paper, we are able
to generate visualizations with {\em correct visual properties}
on querying less than 0.02\% of the data on very large datasets (with
$10^{10}$ tuples), 
giving us a speed-up of over 60$\times$ over other schemes (such as \ir) that
provide similar guarantees,
and 1000$\times$ over the scheme that simply generates the visualization
on the entire database.

{\scriptsize
\bibliographystyle{abbrv}
\bibliography{approxbib}
}

\end{document}